\documentclass[10pt]{article}
\usepackage{feynmf}
\usepackage{epsfig}

\begin{document}

\title{Pion Rescattering in Two-Pion Decay of Heavy Quarkonia}

\author{T.A. L\"ahde$\,$\footnote{talahde@pcu.helsinki.fi}$\:$
and$\:\:$D.O. Riska$\,$\footnote{riska@pcu.helsinki.fi}
\\ \vspace{0.3cm} \\
{\normalsize \it Helsinki Institute of Physics,}\\
{\normalsize \it PL 64 University of Helsinki, 00014 Finland} }

\date{}
\maketitle
\thispagestyle{empty}

\begin{abstract}
The role of pion rescattering in $\pi\pi$ decay of radially excited heavy 
quarkonia modeled in terms of a $Q\pi\pi$ coupling, is investigated 
within the framework of the covariant 
Blankenbecler-Sugar equation. The effects of pion rescattering (or pion 
exchange) are shown to be large, unless the coupling of the two-pion 
system to the heavy quarks is mediated by a fairly light scalar $\sigma$ 
meson, which couples to the gradients of the pion fields. The Hamiltonian 
model for the quarkonium states is formed of linear scalar confining, 
screened one-gluon exchange and instanton induced interaction terms.  
The widths and energy distributions of the basic decays $\psi'\rightarrow 
J/\psi\,\pi\pi$ and $\Upsilon'\rightarrow \Upsilon\,\pi\pi$ are shown to 
be satisfactorily described by this model. The implications of this model 
for the decays of the $\Upsilon(3S)$ state are discussed.
\end{abstract}
\newpage

\section{Introduction}

Two pion decay of radially excited heavy quarkonium ($Q\bar Q$) states
empirically constitutes a significant fraction of their total 
decay widths~\cite{PDG}. Indeed, in the case of the $\psi'$ (or 
$\psi(2S)$), the branching ratio of the $\pi\pi$ decay mode is empirically 
as large as $\sim 50 \%$. Consequently, there has been considerable 
theoretical interest in these decays as the coupling of two pions to heavy 
flavor mesons (or quarks) involves at least two gluons if not a glueball. 
A number of different theoretical approaches for the coupling of two-pions 
to heavy mesons have been proposed, ranging e.g. from effective field theory 
descriptions~\cite{Savage} and directly QCD-motivated 
models~\cite{Shifman} to phenomenological models~\cite{Schwinger}. In 
ref.~\cite{Mannel}, a Lagrangian motivated by chiral perturbation theory 
has been fitted to experiment. The current empirical data on the 
$\pi\pi$ decay of the $\psi'$ and the analogous $\Upsilon'$ (or 
$\Upsilon(2S)$) state, along with a review of the relevant 
literature on the subject is given in ref.~\cite{bbexp}.

Theoretical work on the $\pi\pi$ decays of excited heavy quarkonia has
demonstrated that the empirical energy spectra of the emitted two-pion
system demands that the pions be derivatively coupled to the 
heavy quarkonium states. This is consistent with the role of the pions as 
Goldstone bosons of the spontaneously broken approximate chiral symmetry 
of QCD. Most models~\cite{Savage,Shifman,Schwinger,Mannel} have dealt
with the coupling of two-pions to the heavy meson as a whole rather than 
to its constituent quarks. The satisfactory description obtained
suggests that the decay amplitude $T_{fi}$ at the quark level 
should be a smoothly varying, almost constant, function of the two-pion 
momentum $\vec q$, which is dominated by single-quark mechanisms for two-pion 
emission.

At the quark level, this is, however, not {\it a priori} expected to be the 
case. In the non-relativistic approximation, the amplitude for $\pi\pi$ 
decay of excited $S$-wave states to the ground state through single-quark 
mechanisms is small because of the orthogonality of the quarkonium wave 
functions for different states. In this approximation the
hadronic matrix element does not lead to a constant decay amplitude,
but to one that increases quadratically with the two-pion momentum $\vec 
q$. Furthermore, the pion rescattering or pion exchange term that appears 
naturally as a consequence of the coupling of two-pions to constituent 
quarks is in contrast not suppressed by the orthogonality of the wave 
functions, and may actually be shown to be dominant even in the 
relativistic case.

It is 
shown here that an unrealistically large pion exchange contribution
may be avoided if the $Q\pi\pi$ vertex involves an intermediate, fairly 
light and broad $\sigma$ meson, in line with the phenomenological 
resonance model of ref.~\cite{Schwinger}. An intermediate $\sigma$ meson 
leads to a drastic reduction of the contributions from pion exchange 
mechanisms, while single quark amplitudes are but slightly affected. This 
can be understood qualitatively as a consequence of the exchanged pion 
being off shell. The relativistic treatment of the quark spinors is shown 
to lead to a significant strengthening of the single quark amplitudes
relative to those of pion exchange, and also reproduces the expected 
smooth behavior of the decay amplitude.

The wave functions of the heavy quarkonium states are here obtained as 
solutions to the covariant Blankenbecler-Sugar (BSLT) 
equation~\cite{Blank,Tavkh} with an interaction Hamiltonian, which 
consists of the scalar linear confining + screened one-gluon exchange 
(OGE) model of ref.~\cite{Lahde} and the instanton model employed in 
refs.~\cite{ch1,ch2}. 

A fairly satisfactory description of the empirical $\pi\pi$ decay rates 
and spectra for the transitions $\psi(2S) \rightarrow J/\psi\,\pi\pi$ and 
$\Upsilon(2S) \rightarrow \Upsilon(1S)\pi\pi$ is achieved when the pion 
rescattering contribution is~\mbox{small.} The decay $\Upsilon(3S) 
\rightarrow 
\Upsilon(1S)\pi\pi$ remains something of a puzzle, as the spectrum of the 
two-pion for this transition indicates a dominance of the rescattering 
contribution.

This paper is organized as follows: Section 2 describes the interaction 
Hamiltonian of the $Q\bar Q$ system as employed with the BSLT equation and 
the resulting spectra and wave functions of the $Q\bar Q$ 
states. Section 3 contains the calculation of the $\pi\pi$ width and the 
derivation of the decay amplitude $T_{fi}$ and the hadronic matrix 
elements. In section 4 the results for the $\pi\pi$ decays of the $\psi'$ 
and $\Upsilon'$ states are presented. Section 5 contains a discussion 
of the obtained results for the $\pi\pi$ decays of these states along with 
remarks on the implications of this model for the $\pi\pi$ decays 
of the $\Upsilon(3S)$ state.

\newpage

\section{Interaction Hamiltonian for $Q\bar Q$ Systems}

\subsection{Model Hamiltonian}
\label{ham-sec}

The interaction Hamiltonian used in this paper in conjunction with the 
covariant Blankenbecler-Sugar (BSLT) equation to obtain spectra and 
wave functions for the charmonium ($c\bar c$) and bottomonium ($b\bar b$) 
states may be expressed as 
\begin{equation}
H_{\mathrm {int}} = V_{\mathrm {conf}} + V_{\mathrm {OGE}} + V_{\mathrm 
{inst}}, \label{ham}
\end{equation}
where the different parts denote the confining interaction, one-gluon 
exchange (OGE) and instanton induced interactions respectively. In view of 
the indications given by the radiative M1 transitions in heavy 
quarkonia~\cite{Timo} as well as the M1~\cite{Lahde} and 
pion~\cite{pidec,Goity} decays of the heavy-light mesons, the confining 
interaction is here taken to couple as a Lorentz scalar. The OGE 
interaction motivated by perturbative QCD has vector coupling structure. 
In the nonrelativistic approximation
the interaction potentials in the 
Hamiltonian~(\ref{ham}) may be expressed in the 
form:
\begin{equation}
V_{\mathrm {conf}} = cr
-\frac{c}{r}\frac{M_Q^2+M_{\bar Q}^2}{4M_Q^2M_{\bar Q}^2}\vec S\cdot \vec L
\label{statconf}
\end{equation}
for the scalar confining interaction, where the constant $c$ is 
known~\cite{Timo,Bali,Isgur} to be of the order $\sim 1$ GeV/fm. The 
second term in eq.~(\ref{statconf}) represents the spin-orbit component of 
the confining 
interaction. Note that the antisymmetric spin-orbit interaction will be 
omitted throughout since it vanishes in the case of equal quark and 
antiquark masses. Similarly, the OGE interaction may be expressed 
as~\cite{Davies}
\begin{eqnarray}
V_{\mathrm {OGE}} &=& -\frac{4}{3}\frac{\alpha_s}{r}
+ \frac{2}{3}\frac{\alpha_s}{r^3}
\left(\frac{M_Q^2+M_{\bar Q}^2}{2M_Q^2M_{\bar Q}^2}+\frac{2}{M_QM_{\bar 
Q}}\right) \vec S\cdot \vec L
+\frac{8\pi}{9}\frac{\alpha_s}{M_QM_{\bar Q}}\delta^3(r) \:\:
\vec\sigma_Q\cdot\vec\sigma_{\bar Q} \nonumber \\
&&+\frac{1}{3} \frac{\alpha_s}{M_QM_{\bar Q} r^3} S_{12},
\label{OGEstat}
\end{eqnarray}
where $\alpha_s$ denotes the running coupling constant of perturbative 
QCD, $S_{12}$ is the usual tensor operator $S_{12} = 3(\vec \sigma_Q 
\cdot \hat r)(\vec \sigma_{\bar Q} \cdot \hat r)
-\vec \sigma_Q \cdot \vec \sigma_{\bar Q}$, and $M_Q$, $M_{\bar Q}$ denote 
the heavy quark and antiquark masses, respectively.

The instanton induced 
interaction in systems with heavy quarks has been considered in 
refs.~\cite{ch1,ch2} and consists of a 
spin-independent as well as a $\vec \sigma_Q \cdot \vec \sigma_{\bar Q}$ 
spin-spin interaction term. However, it was also found that only the 
spin-independent term contributes significantly for systems composed of 
two heavy quarks. Thus the instanton contribution to eq.~(\ref{ham}) may 
be expressed as~\cite{ch2}
\begin{equation}
V_{\mathrm{inst}} = - \frac{(\Delta M_Q)^2}{4n} \:\:\delta^3(r),
\label{instpot}
\end{equation}
where $\Delta M_Q$ denotes the mass shift of the heavy constituent quark 
due to the instanton induced interaction. For a charm quark, this is 
expected to be of the order of $\sim 100$ MeV~\cite{ch2}. The parameter 
$n$ represents the instanton density, which is typically assigned values 
around $\sim 1\:\mathrm{fm}^{-4}$.

\subsection{The Blankenbecler-Sugar equation}
\label{blank-sec}

The covariant Blankenbecler-Sugar equation for a quark-antiquark system 
may be expressed as an eigenvalue equation similar to the nonrelativistic 
Schr\"odinger equation:
\begin{equation}
\left(-\frac{\vec \nabla\,^2}{2\mu} + V\right)
\Psi_{\mathrm{nlm}}(\vec r) =
\varepsilon \Psi_{\mathrm{nlm}}(\vec r). \label{diffeq}
\end{equation}
Here $\mu$ denotes the reduced mass of the quark-antiquark system, 
while the eigenvalue $\varepsilon$ is related to the energy $E$ of the 
$Q\bar Q$ state as
\begin{equation}
\varepsilon = \frac{\left[ E^2 - (M_Q+M_{\bar Q})^2 \right ]
\left[ E^2 - (M_Q-M_{\bar Q})^2 \right ]}{8\mu E^2}. \label{EBSLT}
\end{equation}  
For equal quark and antiquark masses, this expression simplifies 
to $(E^2 - 4M_Q^2)/4M_Q$. The relation between the
interaction operator $V$, which in general is 
nonlocal, to the $Q\bar Q$ irreducible 
quasipotential $\cal V$ is (in momentum space) 
\begin{equation}
V(\vec p\,',\vec p\,) = \sqrt{\frac{M_Q+M_{\bar Q}}{W(\vec p\,')}} {\cal 
V} (\vec p\,',\vec p\,) \sqrt{\frac{M_Q+M_{\bar Q}}{W(\vec p\,)}}, 
\label{VBSLT}
\end{equation}
where the function $W$ is defined as $W(\vec p\,) = E_Q(\vec p\,) + 
E_{\bar Q}(\vec p\,)$ 
with $E_Q(\vec p\,) = \sqrt{M_Q^2 + \vec p\,^2}$. In 
the Born approximation the quasipotential $\cal V$ is set equal
to the $Q\bar Q$ 
invariant scattering amplitude $\cal T$, and thus a constructive relation 
to field theory obtains.

Comparison of the BSLT equation to the nonrelativistic Schr\"odinger equation
reveals that the quadratic mass operator employed by the 
former~(\ref{EBSLT}) leads to an effective weakening of the repulsive 
kinetic energy operator. Also, wavefunctions that are solutions to 
eq.~(\ref{diffeq}), allow retention of the conventional quantum mechanical 
operator structure. For 
systems that contain charm quarks the static interaction Hamiltonian 
described above
has but qualitative value, because of the slow convergence of the  
asymptotic expansion in $v/c$
~\cite{Lahde}.
Therefore the potentials in eq.~(\ref{ham}) should be
replaced with versions, which take into account both the
modification~(\ref{VBSLT}) and the relativistic effects associated 
with the Lorentz structure of the interaction, as well as the effects of 
the running coupling $\alpha_s(k^2)$. If the nonlocal effects 
of quadratic and higher order in $\vec P = (\vec p\,' + \vec p\,)/2$ are 
dropped, this may be achieved 
by replacing the 
main term of the OGE interaction by~\cite{Lahde} 
\begin{equation}
V_{\mathrm {OGE}} = - \frac{4}{3}\frac{2}{\pi} 
\int_{0}^{\infty}dk\:j_0(kr) \frac{M_Q}{e_Q}\frac{M_{\bar Q}}{e_{\bar Q}}
\left(\frac{M_Q+M_{\bar Q}}{e_Q+e_{\bar Q}}\right)\alpha_s(k^2),
\label{OGEpot}
\end{equation}
where the factors $e_Q$ and $e_{\bar Q}$ are defined as
\begin{equation}
e_Q=\sqrt{M_Q^2+\frac{k^2}{4}},\quad   
e_{\bar Q}=\sqrt{M_{\bar Q}^2+\frac{k^2}{4}}.
\end{equation}
For the running QCD coupling $\alpha_s(\vec k^2)$, the  
parameterization of ref.~\cite{Stevenson}:
\begin{equation}
\alpha_s(k^2)=\frac{12\pi}{27}\frac{1}
{\ln [(k^2+4m_g^2)/\Lambda_0^2]}.\label{zzz}
\end{equation}
has been employed.
Here $\Lambda_0$ and $m_g$ are parameters that determine the high- and 
low-momentum transfer behavior of $\alpha_s$ respectively, which will be 
determined by a fit to the experimental heavy quarkonium spectra. With 
exception of the effects of the running coupling $\alpha_s$, which lead to 
a strengthening of the long-range part of the OGE interaction, the 
relativistic modifications in eq.~(\ref{OGEpot}) lead to a strong 
suppression of the short-range Coulombic potential. These 
relativistic modifications are however significant only for distances 
$<0.5$ fm. In principle, a relativistic form for the linear scalar 
confining interaction analogous to eq.~(\ref{OGEpot}) may also be 
constructed, but in view of its long-range nature the resulting 
modifications are less significant. A similar modification of the 
spin-orbit potentials associated with the OGE and confining 
interactions gives
\small
\begin{eqnarray}
&&V_{\mathrm {OGE}}^{\mathrm {LS}} = \frac{2}{3 \pi r}\vec S\cdot 
\vec L \int_0^\infty dk\:k\:j_1(kr)\left(\frac{M_Q+M_{\bar Q}}{e_Q+e_{\bar 
Q}}\right)
\left(\frac{e_Q+M_Q}{e_Q}\right)\left(\frac{e_{\bar Q}+M_{\bar Q}}{e_{\bar 
Q}}\right) \alpha_s(k^2) \label{OGEls} \\
&& \left\{\frac{1}{(e_{\bar 
Q}+M_{\bar Q})^2}\left[1-\frac{k^2}{4(e_Q+M_Q)^2}\right]
+\frac{1}{(e_Q+M_Q)^2}\left[1-\frac{k^2}{4(e_{\bar Q}+M_{\bar Q})^2}
\right] + \frac{4}{(e_Q+M_Q)(e_{\bar Q}+M_{\bar Q})}\right\} \nonumber
\end{eqnarray}
\normalsize
for the OGE spin-orbit interaction, and
\begin{equation}
V_{\mathrm {conf}}^{\mathrm {LS}} = - {2\over\pi}{c\over r}\vec S\cdot 
\vec L\int_0^{\infty} dk\:\frac{j_1(kr)}{k}
\left(\frac{M_Q+M_{\bar Q}}{e_Q+e_{\bar Q}}\right)
\left(\frac{1}{e_Q(e_Q+M_Q)}+\frac{1}{e_{\bar Q}(e_{\bar 
Q}+M_{\bar Q})}\right)
\label{Confls}
\end{equation}
for the confinement spin-orbit interaction. Note that eqs.~(\ref{OGEls}) 
and~(\ref{Confls}) are free of singularities 
that have to be treated in first order perturbation theory, particularly 
the $r^{-3}$ singularity of the OGE spin-orbit interaction 
which is an illegal operator in the differential equation~(\ref{diffeq}). 
The modification of the spin-spin delta function term in 
eq.~(\ref{OGEstat}) is obtained as
\begin{equation}
V_{\mathrm {OGE}}^{\mathrm {SS}} = 
\frac{4}{9\pi}\,\vec\sigma_Q\cdot\vec\sigma_{\bar Q}\,  
\int_0^\infty dk\:k^2\:j_0(kr)
\left(\frac{M_Q+M_{\bar Q}}{e_Q+e_{\bar Q}}\right)\frac{\alpha_s(k^2)}{e_Q 
e_{\bar Q}}
\label{OGEss}
\end{equation}
and that of the tensor interaction as
\begin{equation}
V_{\mathrm {OGE}}^{\mathrm {T}} =
\frac{2}{9 \pi} S_{12}\int_0^\infty dk\:k^2\:j_2(kr)
\left(\frac{M_Q+M_{\bar Q}}{e_Q +e_{\bar Q}}\right) 
\frac{\alpha_s(k^2)}{e_Q e_{\bar Q}}. 
\label{OGEt}
\end{equation}
These forms are also free of singularities that have to be treated in 
first order perturbation theory. Note that for $r>0.5$ fm the modified 
potentials asymptotically approach the static forms of 
eqs.~(\ref{statconf}) and~(\ref{OGEstat}), if $\alpha_s$ is taken to be 
constant. In case of the instanton induced interaction, the delta function 
of eq.~(\ref{instpot}) is modified as
\begin{equation}
V_{\mathrm{inst}} = - \frac{(\Delta M_Q)^2}{4n} \:\:
\int_0^\infty dk\:k^2\:j_0(kr)
\left(\frac{M_Q+M_{\bar Q}}{e_Q+e_{\bar Q}}\right)\frac{M_QM_{\bar Q}}{e_Q
e_{\bar Q}},
\label{instpot2}
\end{equation}
leading to a smeared out form of the delta function, which allows for a 
direct numerical treatment of the instanton potential. In the limit of 
very large quark and antiquark masses (the static limit) the above 
equation reduces to the form~(\ref{instpot}).

\begin{table}[h!]
\begin{center}
\begin{tabular}{c||r|c}
Parameter & Current value & Other sources \\ \hline\hline &&\\
$M_b$	& 4885 MeV	& 4870 MeV~\cite{Roberts}	\\
$M_c$	& 1500 MeV	& 1530 MeV~\cite{Roberts}	\\ &&\\
$\Lambda_{\mathrm{QCD}}$ & 260 MeV & 200-300 MeV~\cite{Stevenson} \\
$m_g$   & 290 MeV	& $m_g > \Lambda_{\mathrm{QCD}}$~\cite{Stevenson}\\
$c$	& 890 MeV/fm	& 912 MeV/fm~\cite{Roberts} \\	&&\\
$\frac{(\Delta M_c)^2}{4n}$ & 0.084 $\mathrm{fm}^{-6}$ & $\sim 0.05$ 
$\mathrm{fm}^{-6}$~\cite{ch2} \\ 
$\frac{(\Delta M_b)^2}{4n}$ & 0.004 $\mathrm{fm}^{-6}$ & ? \\ 
\end{tabular}
\caption{Quark masses and coupling constants corresponding to the 
calculated spectra in Fig.~\ref{spektrfig}. The heavy masses 
are close to those used in ref.~\cite{Roberts}, and 
in general in agreement with the values in earlier work. The values of 
$\Lambda_{\mathrm{QCD}}$ and $m_g$ are also in line with the general 
criteria of ref.~\cite{Stevenson}, and the confining interaction strength 
$c$ is also in good agreement with the values of earlier 
calculations~\cite{Roberts,Isgur}. For charmonium, the strength of the 
instanton induced interaction is comparable to the estimate given by 
ref.~\cite{ch2}. } 
\label{partab}
\end{center} 
\end{table}

\newpage

\subsection{Heavy Quarkonium Spectra and Wave functions}
\label{Spec-sec}

\begin{figure}[h!]
\parbox{0.5\textwidth}{\epsfig{file = 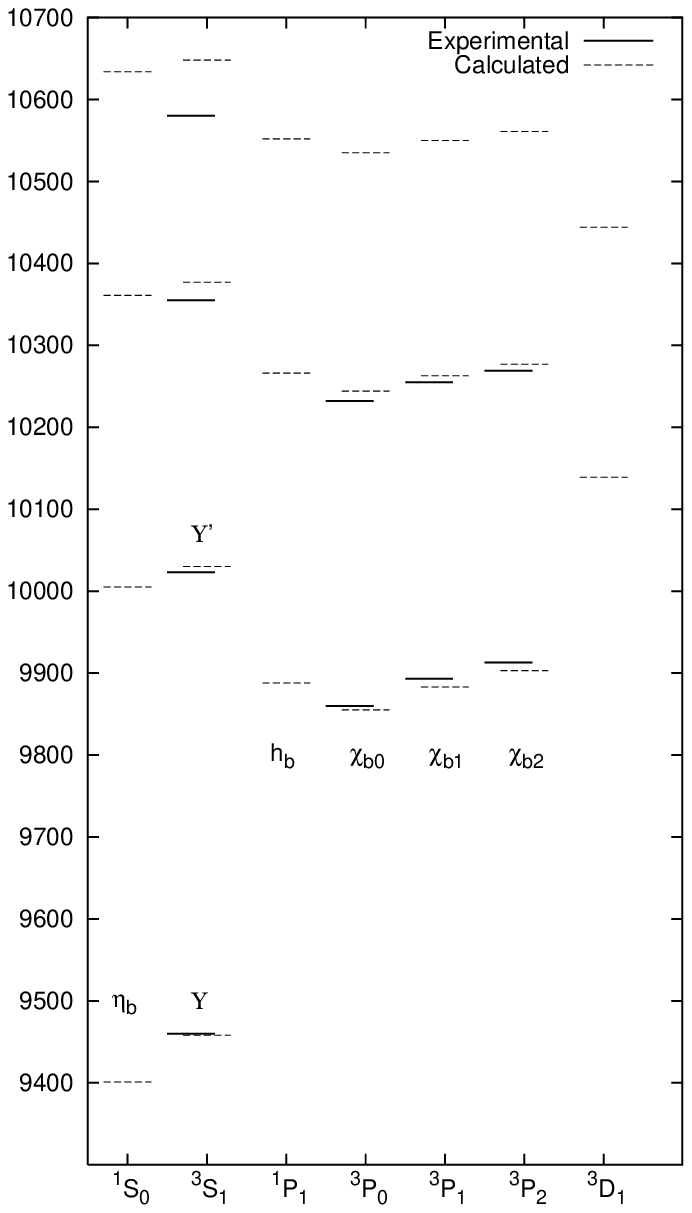}}
\parbox{0.5\textwidth}{\epsfig{file = 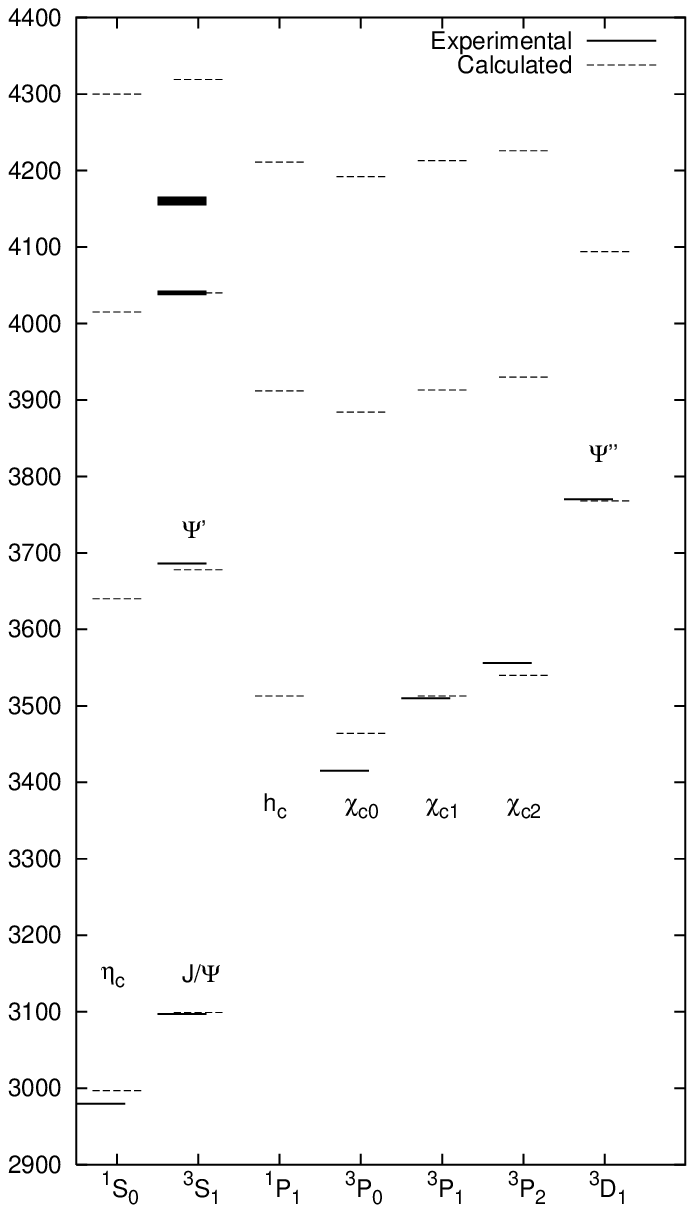}}
\caption{Calculated and experimental Bottomonium($b\bar b$) and 
Charmonium($c\bar c$) spectra. All states are given in MeV, and correspond 
to the data in Table~\ref{spektrtab}. The thickness of the 
lines denoting the experimentally determined states indicate the 
uncertainty in the mass of the state. Note that the identification of the 
$4\,^3S_1$ state in charmonium is uncertain, and may actually be a 
$2\,^3D_1$ state, or a mixture of the two. The threshold for $D\bar D$ 
decay is at $\sim 3750$ MeV, and for $B\bar B$ decay at $\sim 10500$ MeV.} 
\label{spektrfig} 
\end{figure}

The heavy quark masses and the parameters in the quark-antiquark 
interaction operator are determined by a fit to the empirical 
charmonium and bottomonium spectra, which was achieved by
numerical solution of eq.~(\ref{diffeq}) for the quarkonium
states with the Runge-Kutta-Nystr\"om (RKN) algorithm.
The resulting spectra are presented in Fig.~\ref{spektrfig} and the 
corresponding parameters of the $Q\bar Q$ interaction operator are given 
in Table~\ref{partab}.

\newpage

\begin{table}[h!]
\parbox{0.6\textwidth}{
\begin{tabular}{c||c|c|c|c} 
$n\,^{2S+1}L_J$ & $b\bar b$ & Exp($b\bar b$) & $c\bar c$ & Exp($c\bar c$) 
\\ \hline\hline
&&&&\\
$1\,^1S_0$ & 9401  &  --  & 2997 & $2980\pm 1.8$  \\
$2\,^1S_0$ & 10005 &  --  & 3640 &  --	\\
$3\,^1S_0$ & 10361 &  --  & 4015 &  --	\\
$4\,^1S_0$ & 10634 &  --  & 4300 &  --	\\ 
&&&&\\
$1\,^3S_1$ & 9458  & 9460  & 3099 & 3097 \\
$2\,^3S_1$ & 10030 & 10023 & 3678 & 3686 \\
$3\,^3S_1$ & 10377 & 10355 & 4040 & $4040\pm 10$ \\
$4\,^3S_1$ & 10648 & 10580 & 4319 & $4159\pm 20$ ? \\
&&&&\\
$1\,^1P_1$ & 9888  &  --  & 3513 &  --	\\
$2\,^1P_1$ & 10266 &  --  & 3912 &  --	\\
$3\,^1P_1$ & 10552 &  --  & 4211 &  --	\\
&&&&\\
$1\,^3P_0$ & 9855  & 9860  & 3464 & 3415 \\
$2\,^3P_0$ & 10244 & 10232 & 3884 &  --	\\
$3\,^3P_0$ & 10535 &  --   & 4192 &  --	\\
&&&&\\
$1\,^3P_1$ & 9883  & 9893  & 3513 & 3511 \\
$2\,^3P_1$ & 10263 & 10255 & 3913 &  --	\\
$3\,^3P_1$ & 10550 &  --   & 4213 &  --	\\
&&&&\\
$1\,^3P_2$ & 9903  & 9913  & 3540 & 3556 \\
$2\,^3P_2$ & 10277 & 10269 & 3930 &  --	\\
$3\,^3P_2$ & 10561 &  --   & 4226 &  --	\\
&&&&\\
$1\,^3D_1$ & 10139 &  --  & 3768 & $3770\pm 2.5$ \\
$2\,^3D_1$ & 10444 &  --  & 4094 &  -- ? \\
\end{tabular}}
\parbox{0.4\textwidth}{
\caption{Calculated and experimental charmonium and bottomonium states 
rounded to the nearest MeV. The experimental states correspond to the 
values reported by 
ref.~\cite{PDG}. The states are classified according to excitation number 
$n$, total spin $S$, total orbital angular momentum $L$ and total angular 
momentum $J$. These values are plotted in Fig.~\ref{spektrfig}. Experimental 
uncertainties are indicated only where they are appreciable.
\label{spektrtab}}}
\end{table}

The calculated states listed in Table~\ref{spektrtab} are, for the most 
part, in good agreement with the experimental results. The agreement is 
similar to that achieved in ref.~\cite{Roberts} as well as in
ref.~\cite{Bali}, where the heavy quark potential was determined by 
numerical lattice QCD calculation. The threshold for 
$D\bar D$ fragmentation in charmonium 
lies at $\sim 3750$ MeV, while that for $B\bar B$ 
fragmentation in bottomonium lies at $\sim 10500$ MeV. 
The states 
that lie above the threshold for strong decay
are less well described by the present model, a feature that is shared 
with other similar models, e.g. ref.~\cite{Roberts}. 
Below threshold, the most significant discrepancy is 
the too small hyperfine splittings obtained for the $L=1$ states in 
charmonium. This problem appears to be common for all calculations based 
on interaction Hamiltonians formed of OGE + scalar 
confinement components~\cite{Bali,Isgur,Roberts}. Here this 
problem may be traced to the weakness of the OGE tensor interaction as 
given by eqs.~(\ref{OGEstat}) and~(\ref{OGEt}). 

The coupling constants used in the interaction Hamiltonian were optimized 
by fits to the empirical charmonium and bottomonium spectra, and are 
listed in Table~\ref{partab}. In order to accommodate universal parameters 
for the OGE and linear confining interactions, perfect agreement with 
experiment had to be sacrificed in the bottomonium sector. The resulting 
central (spin-independent) components of the quark-antiquark potential 
turn out to be quite well described by a Cornell form $V(r) = -a/r + cr + 
b$, with a Coulombic part that is somewhat stronger for the charmonium 
system.

\newpage

\section{The Model for $\pi\pi$ decay}

\subsection{Width for $\pi\pi$ decay}
\label{dec-sec}

The general expression for the two pion decay width of an excited $Q\bar 
Q$ meson may be written in the form
\begin{equation}
\Gamma = (2\pi)^4\int\frac{d^3k_a}{(2\pi)^3}\frac{d^3k_b}{(2\pi)^3}  
\frac{d^3P_f}{(2\pi)^3} \frac{M_fM_i}{E_fE_i} \frac{|T_{fi}|^2}
{4\omega_a\omega_b} \delta^{(4)}(P_f + k_a
+ k_b -P_i).
\end{equation}
Here $k_a$ and $k_b$ denote the four-momenta of the two emitted pions,
$P_i$ and $P_f$ those of the initial and final state $Q\bar Q$ 
mesons while $\omega_a$ and $\omega_b$ denote the energies of the  
emitted pions respectively. The factors $M/E$ are normalization factors 
for the heavy meson states similar to those employed in ref.~\cite{Goity}.
Since the constituents of the $Q\bar Q$ mesons form bound
states, their normalization factors are included in the spinors $\bar
u(p')$ and $u(p)$ in eq.~(\ref{ampl}). In the laboratory frame
$P_i^0 = E_i = M_i$. By introducing the variables $\vec Q = (\vec k_b - 
\vec k_a)/2$ and $\vec q = \vec k_b + \vec k_a$, the decay
width expression may be rewritten as
\begin{equation}
\Gamma = \int\frac{d^3q d^3Q}{(2\pi)^5} \frac{M_f}{E_f}\frac{|T_{fi}|^2}
{4\omega_a\omega_b}\:\:
\delta \left(\sqrt{q^2 + M_f^2} + \omega_a + \omega_b - M_i\right).
\end{equation}
Here the pion energy factors are defined as $\omega_a = \sqrt{m_\pi^2 + 
(\vec q/2 -\vec Q)^2}$ and $\omega_b = \sqrt{m_\pi^2 + (\vec q/2 +\vec 
Q)^2}$ respectively. The remaining delta function determines the
variable $\vec Q$, so that finally the expression for the differential
width becomes
\begin{equation}
\frac{d\Gamma}{d\Omega_q} = \frac{1}{4}\frac{1}{(2\pi)^4} \int_0^{q_f}
dq\:q^2 \int_{-1}^1 dz\:\frac{Q_f^2(q,z)}{\omega_a(q,z)\left(Q_f +
\frac{qz}{2}\right) + \omega_b(q,z)
\left(Q_f - \frac{qz}{2}\right)}\:\frac{M_f}{E_f(q)}\:|T_{fi}|^2.
\label{dec}
\end{equation}
In eq.~(\ref{dec}), the variable $z$
is defined by $\vec Q\cdot\vec q = Qqz$, and $E_f$ denotes the energy 
of the final state $Q\bar Q$ meson and is given by $E_f = \sqrt{q^2 + 
M_f^2}$. With this notation the pion
energies $\omega_a$ and $\omega_b$ are given by the expressions 
\begin{equation}
\omega_a = \sqrt{m_{\pi}^2 + Q_f^2 + q^2/4 - Q_fqz}\quad \mathrm{and}\quad
\omega_b = \sqrt{m_{\pi}^2 + Q_f^2 + q^2/4 + Q_fqz}, 
\end{equation}
where the fixed variable $Q_f$ is given by
\begin{equation}
Q_f^2 = \frac{(E_f - M_i)^4 - (4m_{\pi}^2 + q^2)(E_f - M_i)^2}{4(E_f -
M_i)^2 - 4q^2z^2}.
\end{equation}
In the expressions above, $m_\pi$ denotes the pion mass. Since different 
charge states are considered in this paper, slightly different pion masses 
will be used for decays that involve charged or neutral pions. The 
integration limit $q_f$ corresponds to the maximal  
momentum of any one of the final state particles, e.g. the final state 
$Q\bar Q$ meson. Thus $q_f$ corresponds to the q-value of a decay of the 
form $A'\rightarrow AX$, where $A'$ and $A$ are the appropriate $Q\bar Q$ 
meson states and $X$ is a particle with mass $M_X = 2m_\pi$. The 
appropriate values of $q_f$ for the different decays are listed in 
Table~\ref{restab}.

The above equations are ideally suited for computation of the $\pi\pi$ 
decay width using a hadronic matrix element $T_{fi}$. However, 
experimental data is usually presented~\cite{bbexp} in terms of a 
dimensionless variable $x$, which is defined as
\begin{equation}
x = \frac{m_{\pi\pi} - 2m_\pi}{\Delta M}.
\label{xvar}
\end{equation}
Here $m_{\pi\pi}$ denotes the invariant mass $\sqrt{s_{\pi\pi}}$ of the 
two-pion system, and $\Delta M = M_i - M_f - 2m_{\pi}$. With these 
definitions, the variable $x$ is always between 0 and 1. The relation 
between $q$ and $m_{\pi\pi}$ may then be obtained as~\cite{PDG}
\begin{equation}
|\:\vec q\:| = \frac{\sqrt{\left(M_i^2 - (M_f + m_{\pi\pi})^2\right)
\left(M_i^2 - (M_f - m_{\pi\pi})^2\right)}}{2M_i}.
\end{equation}
From this relation, the Jacobian of transformation may be 
obtained as
\begin{equation}
\frac{dq}{dx} = \frac{\Delta M}{4M_i^2q(m_{\pi\pi})}
\left\{\left[M_i^2 - (M_f + m_{\pi\pi})^2\right]
(M_f - m_{\pi\pi}) - \left[M_i^2 - (M_f - m_{\pi\pi})^2\right]
(M_f + m_{\pi\pi})\right\}.
\end{equation}
Finally the total decay width may be calculated as
\begin{equation}
\Gamma = \int_0^{q_f} dq \frac{d\Gamma}{dq} = - \int_0^1 dx 
\frac{d\Gamma}{dq} \frac{dq}{dx},
\label{dwidth}
\end{equation}
where the minus sign appears because the integration limits have been 
reversed to take into account the fact that $q=q_f$ corresponds to $x=0$. 
In eq.~(\ref{dwidth}), the latter form turns out to be the most convenient 
since experiments generally present the measurements of the $\pi\pi$ 
energy distribution using dimensionless quantities, e.g. by plotting 
$1/\Gamma (d\Gamma/dx)$ versus $x$. In the following section, the model 
for the $\pi\pi$ decay amplitude is presented.

\subsection{Amplitude for $\pi\pi$ decay}

There are several different models for the coupling of two-pions to 
heavy constituent quarks. The emission of two-pions from a heavy flavor 
quark is mediated by two or more gluons or a glueball. For small momenta 
of the two-pion system this
coupling may be approximated by a point coupling, but at larger momenta 
the strong interaction between the pions has to be taken into account, 
e.g. approximately through a broad scalar meson resonance. 

If the coupling of the pions to the constituent quark does not 
involve derivatives of the pion field, agreement with experiment is 
excluded for the pion invariant mass distributions in the decays 
$\Upsilon'\rightarrow \Upsilon\,\pi\pi$ and $\psi'\rightarrow 
J/\psi\,\pi\pi$~\cite{bbexp,Schwinger}. Derivative couplings for the pions 
are also consistent with the role of pions as Goldstone bosons of the 
spontaneously broken approximate chiral 
symmetry of QCD. The model for the $Q\pi\pi$ interaction Lagrangian is 
therefore expected to have the form
\begin{equation}
{\mathcal L}_{Q\pi\pi} = 4\pi\lambda\:\bar\psi_Q 
(\partial_\mu\vec\phi_\pi)\cdot 
(\partial_\mu\vec\phi_\pi)\psi_Q,
\label{Q2pi}
\end{equation}
where $\lambda$ is a coupling constant of dimension~$[\mathrm{MeV}]^{-3}$ 
and $\psi_Q$,$\bar\psi_Q$ denote the heavy quark spinors. 

\begin{figure}
\begin{center}
\begin{tabular}{c c}
\begin{fmffile}{2pisq1}
\begin{fmfgraph*}(130,100) \fmfpen{thin}
\fmfleft{i1,i2}
\fmfright{o1,o2}
\fmftop{o3,o4}
\fmf{fermion}{i2,v1,o2}
\fmf{fermion}{o1,i1}
\fmf{dashes}{v1,o3}
\fmf{dashes}{v1,o4}
\fmfblob{.15w}{v1}
\fmflabel{$Q$}{i2} \fmflabel{$\bar Q$}{i1}
\fmflabel{$k_b$}{o3} \fmflabel{$k_a$}{o4}
\fmfforce{(.1w,.75h)}{i2}
\fmfforce{(.1w,.25h)}{i1}
\fmfforce{(.9w,.75h)}{o2}
\fmfforce{(.9w,.25h)}{o1}
\fmfforce{(.5w,.75h)}{v1}
\fmfforce{(.9w,h)}{o3}
\fmfforce{(.9w,.5h)}{o4}
\end{fmfgraph*}
\end{fmffile}
&
\begin{fmffile}{2pisq2}
\begin{fmfgraph*}(130,100) \fmfpen{thin}
\fmfleft{i1,i2}
\fmfright{o1,o2}
\fmftop{o3,o4}
\fmf{fermion}{o1,v1,i1}
\fmf{fermion}{i2,o2}
\fmf{dashes}{v1,o3}
\fmf{dashes}{v1,o4}
\fmfblob{.15w}{v1}
\fmflabel{$Q$}{i2} \fmflabel{$\bar Q$}{i1}
\fmflabel{$k_b$}{o3} \fmflabel{$k_a$}{o4}
\fmfforce{(.1w,.75h)}{i2}
\fmfforce{(.1w,.25h)}{i1}
\fmfforce{(.9w,.75h)}{o2}
\fmfforce{(.9w,.25h)}{o1}
\fmfforce{(.5w,.25h)}{v1}
\fmfforce{(.9w,0)}{o3}
\fmfforce{(.9w,.5h)}{o4}
\end{fmfgraph*}
\end{fmffile}
\\ &\\
\begin{fmffile}{2piex1}
\begin{fmfgraph*}(130,100) \fmfpen{thin}
\fmfleft{i1,i2}
\fmfright{o1,o2}
\fmftop{o3,o4}
\fmf{fermion}{i2,v1,o2}
\fmf{fermion}{o1,v2,i1}
\fmf{dashes}{v1,o3}
\fmf{dashes,label=$k$}{v1,v2}
\fmf{dashes}{v2,o4}
\fmfblob{.15w}{v1}
\fmfblob{.15w}{v2}
\fmflabel{$Q$}{i2} \fmflabel{$\bar Q$}{i1}
\fmflabel{$k_a$}{o3} \fmflabel{$k_b$}{o4}
\fmfforce{(.1w,.75h)}{i2}
\fmfforce{(.1w,.25h)}{i1}
\fmfforce{(.9w,.75h)}{o2}
\fmfforce{(.9w,.25h)}{o1}
\fmfforce{(.5w,.75h)}{v1}
\fmfforce{(.5w,.25h)}{v2}
\fmfforce{(.9w,.6h)}{o3}
\fmfforce{(.9w,.4h)}{o4}
\end{fmfgraph*}
\end{fmffile}
&
\begin{fmffile}{2piex2}
\begin{fmfgraph*}(130,100) \fmfpen{thin}
\fmfleft{i1,i2}
\fmfright{o1,o2}
\fmftop{o3,o4}
\fmf{fermion}{i2,v1,o2}
\fmf{fermion}{o1,v2,i1}
\fmf{dashes}{v1,o3}
\fmf{dashes,label=$k$}{v1,v2}
\fmf{dashes}{v2,o4}
\fmfblob{.15w}{v1}
\fmfblob{.15w}{v2}
\fmflabel{$Q$}{i2} \fmflabel{$\bar Q$}{i1}
\fmflabel{$k_b$}{o3} \fmflabel{$k_a$}{o4}
\fmfforce{(.1w,.75h)}{i2}
\fmfforce{(.1w,.25h)}{i1}
\fmfforce{(.9w,.75h)}{o2}
\fmfforce{(.9w,.25h)}{o1}
\fmfforce{(.5w,.75h)}{v1}
\fmfforce{(.5w,.25h)}{v2}
\fmfforce{(.9w,.4h)}{o3}
\fmfforce{(.9w,.6h)}{o4}
\end{fmfgraph*}
\end{fmffile} \\
\end{tabular}
\caption{Feynman diagrams for the emission of two-pions by heavy 
constituent quarks. The two upper diagrams correspond to the single-quark 
amplitudes $T_Q$ and $T_{\bar Q}$ respectively, while the two lower 
diagrams describe the pion exchange (or pion rescattering) amplitude 
$T_{\mathrm{ex}}$ and the crossed amplitude $T_{\mathrm{exc}}$.}
\label{Feynfig}
\end{center}
\end{figure}
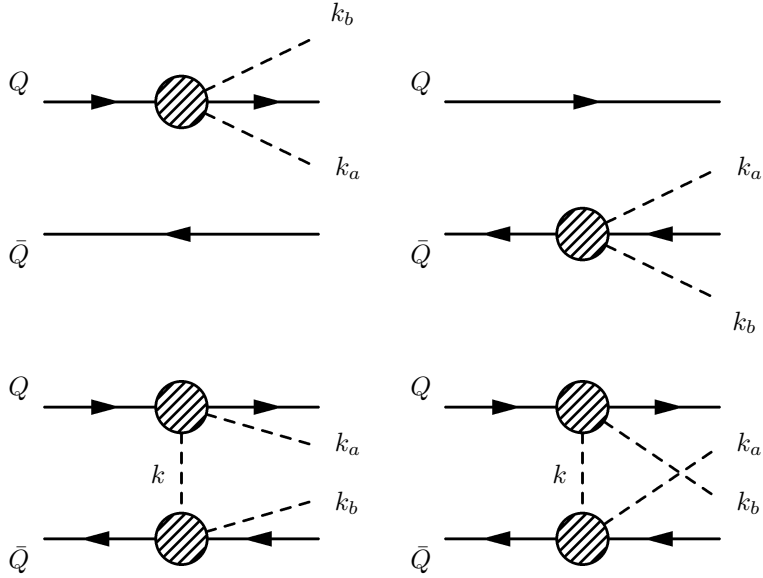

Since the above Lagrangian 
couples the two-pions directly to the constituent quarks of the heavy 
meson, pion exchange mechanisms describing rescattering of the 
emitted pions appear as a natural consequence, in addition to the 
single-quark mechanisms. The relevant Feynman diagrams are shown in 
Fig.~\ref{Feynfig}. 
The amplitude $T_{fi}$ in eq.~(\ref{dec}) may thus be 
expressed in the form
\begin{equation}
T = T_Q + T_{\bar Q} + T_{\mathrm{ex}} + T_{\mathrm{exc}},
\label{ampl}
\end{equation}
where the amplitudes $T_Q$ and $T_{\bar Q}$ are single quark amplitudes 
describing two-pion emission from the quark and antiquark respectively. 
For the equal mass quarkonia considered here, these two amplitudes are 
identical. The amplitudes $T_{\mathrm{ex}}$ and 
$T_{\mathrm{exc}}$ describe diagrams where a pion is exchanged between the 
quark and antiquark during the decay process. $T_{\mathrm{exc}}$ differs 
from $T_{\mathrm{ex}}$ by an interchange of the emitted pions, which will 
be shown to make only a very small difference. The isospin 
dependence of the coupling~(\ref{Q2pi}) implies that

\begin{equation}
|T|^2_{\pi\pi} = 2|T|^2_{\pi^+\pi^-} + |T|^2_{\pi^0\pi^0}.
\end{equation}
As a consequence the width for decay to charged pions should be 
twice that for decay to neutral pions. This is in fair agreement with what
is found experimentally~\cite{PDG,bbexp}. Averaging over initial 
spin states and summing over final spin states introduces nothing further, 
since only the parts of the amplitudes in Fig.~\ref{Feynfig} that are spin 
independent contribute significantly to the decay widths. Treating the 
Feynman diagrams of Fig.~\ref{Feynfig} similarly as the 
pointlike Weinberg-Tomozawa 
interaction for light constituent quarks~\cite{tl2pi}, then the following 
expression describing the single quark diagrams are obtained:
\begin{equation}
T_{\mathrm{1q}} = T_Q + T_{\bar Q} = -2\cdot 8\pi \lambda\: k_a\cdot k_b\:
{\mathcal M}_{\mathrm{1q}},
\label{1q}
\end{equation}
where $k_a$ and $k_b$ are the four-momenta of the emitted pions, and 
${\mathcal M}_{\mathrm{1q}}$ is a matrix element involving the initial and 
final state wave functions and quark spinors. Similarly, the amplitude 
corresponding to the pion-exchange diagrams may be obtained (in momentum 
space) as
\begin{equation}
T_{\mathrm{2q}} = T_{\mathrm{ex}} + T_{\mathrm{exc}} = 
-2\cdot (8\pi \lambda)^2 \frac{k_a\cdot k\:k_b\cdot k}{k^2 + m_\pi^2}.
\label{2q}
\end{equation}
In eq.~(\ref{2q}), the expression $k_a\cdot k\:k_b\cdot k$ may be 
written in the form
\begin{equation}
k_a\cdot k\:k_b\cdot k = \frac{\vec k_a\cdot\vec k_b\:\vec k\,^2}{3} 
+\left[\vec k_a\cdot\vec k\:\vec k_b\cdot\vec k - \frac{ 
\vec k_a\cdot\vec k_b\:\vec k\,^2}{3} \right]
- \vec k_a\cdot\vec k\:\omega_b k_0 - \vec k_b\cdot\vec 
k\:\omega_a k_0 + \omega_a\omega_b k_0^2.
\label{terms}
\end{equation}
If only the $S$-wave pion production terms (i.e. the
first term on the r.h.s. of the above equation and the term 
proportional to $k_0^2$) are retained, eq.~(\ref{2q}) may be expressed 
as
\begin{equation}
T_{\mathrm{2q}} = -2\cdot (8\pi \lambda)^2 
\left\{\frac{1}{3}\left(\frac{\vec q\,^2}{4}-Q_f^2\right)
\left[1-\frac{A^2}{k\,^2 + A^2}\right]
+ \frac{\omega_a\omega_b}{4}(\omega_a - \omega_b)^2\frac{1}{k\,^2 + 
A^2}\right\},
\label{2q2}
\end{equation}
where $A = \sqrt{m_\pi^2 - k_0^2}$. Here $k_0$ denotes the time component 
of $k$. In eq.~(\ref{2q2}), the left-hand term 
will, upon Fourier transformation, contain a delta function. Thus, although 
eq.~(\ref{2q}) is formally of second order in $\lambda$, it may actually 
be numerically significantly larger than the single quark 
amplitude given by eq.~(\ref{1q}), if bare vertices are assumed for the 
$Q\pi\pi$ coupling. Since the kinematical factors associated with the 
pion exchange amplitude differ from that of the single quark
amplitude, dominant pion exchange does not agree with earlier 
studies~\cite{Shifman,Schwinger,Mannel}, which achieved
fair agreement with experimental spectra with effectively single-quark 
amplitudes alone. The indications both from earlier studies of the 
$2S\rightarrow 1S\,\pi\pi$ decays and experiment~\cite{bbexp} are 
therefore 
that the contribution from pion-exchange diagrams should be small. The 
task here is therefore to attempt an explanation for this  
experimental feature.

The effect of the strong interaction between two pions in the
$S-$wave $\pi\pi$ may be approximately accounted for
by inclusion of an intermediate scalar meson ($\sigma$ or glueball) 
resonance in the vertex. This is brought about by modification of 
the coupling constant $\lambda$ with a relativistic 
Breit-Wigner-like scalar meson propagator:
\begin{equation}
\lambda \rightarrow \lambda\left(\frac{M_\sigma^2 + \Gamma_\sigma^2/4} 
{M_\sigma^2 + q^2 + \Gamma_\sigma^2/4}\right).
\label{sigmares}
\end{equation}
Here $M_\sigma$ denotes the pole position $m_\sigma-i\Gamma_\sigma/2$ 
of the effective scalar meson and $q$ the four-momentum of the scalar 
meson ($\sigma$) resonance. The $\sigma$ resonance appears by infinite 
iteration of the four-pion vertex in the isospin 0 spin 0 channel. 
Therefore, as pointed out in ref.~\cite{Schwinger}, it is natural to 
describe the strongly interacting $\pi\pi$ state by a broad $\sigma$ pole 
rather than by the driving term (4-pion vertex) alone. 

If the variables 
defined in section~\ref{dec-sec} are employed, the modifications affecting 
the single-quark amplitude~(\ref{1q}) are relatively straightforward. 
However, the pion-exchange amplitude~(\ref{2q}) becomes much more 
complicated, which is illustrated in Fig.~\ref{sigmafig}:

\begin{figure}[h!]
\parbox{0.48\textwidth}{
\caption{Pion exchange diagram for $\pi\pi$ decay with $Q\pi\pi$ vertices 
modeled with intermediate heavy $\sigma$ mesons. Here the $\sigma$ meson 
4-momenta are defined as $k_1 = -k -k_a$ and $k_2 = k - k_b$. Note that 
just as in Fig.~\ref{Feynfig}, there is also a diagram with $k_a$ and 
$k_b$ interchanged.}
\label{sigmafig}}
\parbox{0.48\textwidth}{\begin{center}
\begin{fmffile}{sigmaex}
\begin{fmfgraph*}(150,150) \fmfpen{thin}
\fmfcmd{%
 vardef port (expr t, p) = 
  (direction t of p rotated 90)
   / abs (direction t of p)
 enddef;}
\fmfcmd{%
 vardef portpath (expr a, b, p) =
  save l; numeric l; l = length p;
  for t=0 step 0.1 until l+0.05:
   if t>0: .. fi point t of p
    shifted ((a+b*sind(180t/l))*port(t,p))
  endfor
  if cycle p: .. cycle fi
 enddef;}
\fmfcmd{%
 style_def brown_muck expr p =
  shadedraw(portpath(thick/2,2thick,p)
   ..reverse(portpath(-thick/2,-2thick,p))
   ..cycle)
 enddef;}
\fmfleft{i1,i2}
\fmfright{o1,o2}
\fmftop{o3,o4}
\fmf{fermion}{i2,v1,o2}
\fmf{fermion}{o1,v2,i1}
\fmf{dashes}{v3,o3}
\fmf{dashes,label=$\pi(k)$}{v3,v4}
\fmf{dashes}{v4,o4}
\fmfdot{v1,v2,v3,v4}
\fmf{brown_muck,lab.s=right,lab.d=4thick,lab=$\sigma(k_1)$}{v1,v3}
\fmf{brown_muck,lab.s=right,lab.d=4thick,lab=$\sigma(k_2)$}{v2,v4}
\fmflabel{$Q$}{i2} \fmflabel{$\bar Q$}{i1}
\fmflabel{$\pi(k_a)$}{o3} \fmflabel{$\pi(k_b)$}{o4}
\fmfforce{(.1w,.95h)}{i2}
\fmfforce{(.1w,.05h)}{i1}
\fmfforce{(.9w,.95h)}{o2}
\fmfforce{(.9w,.05h)}{o1}
\fmfforce{(.5w,.95h)}{v1}
\fmfforce{(.5w,.05h)}{v2}
\fmfforce{(.6w,.70h)}{v3}
\fmfforce{(.6w,.30h)}{v4}
\fmfforce{(.9w,.6h)}{o3}
\fmfforce{(.9w,.4h)}{o4}
\end{fmfgraph*}
\end{fmffile}\end{center}}
\end{figure}
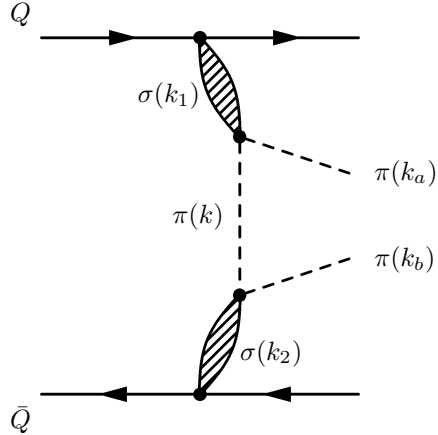

When the modification~(\ref{sigmares}) is taken into account, the 
expression for the single-quark amplitude~(\ref{1q}) becomes
\begin{equation}
T_{\mathrm{1q}} =
-2\cdot 8\pi \lambda \left(\frac{M_\sigma^2 + \Gamma_\sigma^2/4} 
{M_\sigma^2 + q^2 + \Gamma_\sigma^2/4}\right)\left[m_\pi^2 - 
\frac{1}{2}\left((\omega_a + \omega_b)^2 - \vec q\,^2\right)\right]
{\mathcal M}_{\mathrm{1q}}.
\label{1q2}
\end{equation}
In case of the above single-quark amplitude, the nonrelativistic 
approximation for the matrix element ${\mathcal M}_{\mathrm{1q}}$ cannot 
be expected to be reliable, even though the quark masses are large. A 
relativistic form for the matrix element ${\mathcal M}_{\mathrm{1q}}$ may 
be obtained as
\begin{eqnarray}
{\mathcal M}_{\mathrm{1q}}^{\mathrm{rel}} &=& \frac{1}{\pi}
\int_0^\infty dr'\,r'\,u_f(r')\,\int_0^\infty dr\,r\,u_i(r)
\int_0^\infty dP\,P^2 \int_{-1}^1 dv\:\alpha(P,v,q) \nonumber \\
&& j_0\left(r'\sqrt{P^2 + \frac{q^2}{16} - \frac{Pqv}{2}}\:\right)\:
j_0\left(r\sqrt{P^2 + \frac{q^2}{16} + \frac{Pqv}{2}}\:\right),
\label{relmat}
\end{eqnarray}
where $\alpha(P,v,q)$ is a factor which includes the quark spinors in the 
coupling~(\ref{Q2pi}),
\begin{equation}
\alpha(P,v,q) = \sqrt{\frac{(E+M_Q)(E'+M_Q)}{4EE'}}
\left(1-\frac{P^2 - \vec q\,^2/4}{(E'+M_Q)(E+M_Q)}\right),
\label{alpha}
\end{equation}
where the quark energy factors are defined as
\begin{equation}
E = \sqrt{M_Q^2 + P^2 + Pqv + \vec q\,^2/4}, \quad\quad
E' = \sqrt{M_Q^2 + P^2 - Pqv + \vec q\,^2/4}.
\end{equation}
If the nonrelativistic limit is nevertheless employed, eq.~(\ref{relmat}) 
reduces to
\begin{equation}
{\mathcal M}_{\mathrm{1q}}^{\mathrm{nr}} = \int_0^\infty dr\, 
u_f(r)u_i(r)\: j_0\left(\frac{qr}{2}\right).
\label{nrmat}
\end{equation}
In principle the quark spinors in the coupling~(\ref{Q2pi}) also contain 
a spin dependent part that is proportional to $\vec q\times\vec P$. For 
the present purposes that contribution turns out to be very tiny and may 
be safely neglected. It will be shown in the next section that the 
relativistic modifications to the single-quark matrix element lead to a 
significant increase of the single quark amplitudes for $\pi\pi$ decay and 
are important for obtaining agreement with the experimental results.

The pion exchange amplitude corresponding to Fig.~\ref{sigmafig} may 
be expressed as
\begin{equation}
T_{\mathrm{ex}} = -(8\pi \lambda)^2(M_\sigma^2 + \Gamma_\sigma^2/4)^2  
\frac{k_a\cdot k\:k_b\cdot k}{(k_1^2+M_\sigma^2 + \Gamma_\sigma^2/4)
(k^2+m_\pi^2)(k_2^2+M_\sigma^2 + \Gamma_\sigma^2/4)},
\label{3prop}
\end{equation}
where the momenta are defined as in Fig.~\ref{sigmafig}. As the above 
expression contains a triple propagator it turns out to be convenient to 
treat it in an approximate fashion. If one makes the physically reasonable 
assumption that the 3-momenta of the $\sigma$ mesons are about equal in 
magnitude to that of the exchanged pion, and that the emitted pions carry 
away most of the energy while the exchanged pion takes most of the 
3-momentum, then one arrives at the approximations $\vec k_1 \approx -\vec 
k$, $\vec k_2 \approx \vec k$, $|k_1^0| \approx |k_2^0| 
\approx (\omega_a+\omega_b)/2$ and $k_0 \approx (\omega_b-\omega_a)/2$, 
which allow for a simpler treatment of the pion-exchange 
amplitudes. Taking also into account the crossed 
pion exchange term which gives an extra factor of 2, one  arrives at 
the expression
\begin{eqnarray}
T_{\mathrm{2q}} &=& -2\cdot (8\pi \lambda)^2
\left\{\frac{1}{3}\left(\frac{\vec q\,^2}{4}-Q_f^2\right) 
\left[{\mathcal M_{\mathrm e1}} - A^2({\mathcal M_{\mathrm e2}}-{\mathcal 
M_{\mathrm e3}})\right]
+\left(\frac{\vec q\,^2z^2}{4} - \frac{2}{3}Q_f^2 - \frac{\vec 
q\,^2}{12}\right){\mathcal M_{\mathrm e4}} \right.\nonumber \\
&&\left.+\frac{\omega_a\omega_b}{4}(\omega_a - \omega_b)^2
({\mathcal M_{\mathrm e2}}-{\mathcal M_{\mathrm e3}})\right\},
\label{2q3}
\end{eqnarray}  
which replaces eq.~(\ref{2q2}). Note that in eq.~(\ref{2q3}), the term 
proportional to the matrix element $\mathcal M_{\mathrm e4}$ originates 
from the term in square brackets in eq.~(\ref{terms}), which was not 
included in eq.~(\ref{2q2}). The matrix elements in eq.~(\ref{2q3}) may, 
in the non-relativistic approximation, be expressed as
\begin{eqnarray}
{\mathcal M_{\mathrm e1}} &=& \int_0^\infty dr\,u_f(r)u_i(r)\,j_0(Q_fr)\: 
\frac{1}{4\pi}(M_\sigma^2 + \Gamma_\sigma^2/4)^2
\left(\frac{e^{-Xr}}{2X}\right), \\
{\mathcal M_{\mathrm e2}} &=& \int_0^\infty dr\,u_f(r)u_i(r)\,j_0(Q_fr)\:
\frac{1}{4\pi}\frac{(M_\sigma^2 + \Gamma_\sigma^2/4)^2}
{(X^2-A^2)^2} A\,Y_0(Ar), \label{M2}\\
{\mathcal M_{\mathrm e3}} &=& \int_0^\infty dr\,u_f(r)u_i(r)\,j_0(Q_fr)\:
\frac{1}{4\pi}\frac{(M_\sigma^2 + \Gamma_\sigma^2/4)^2}
{(X^2-A^2)^2}\left(X\,Y_0(Xr) + \frac{(X^2-A^2)}{2X}\,e^{-Xr}\right), \\
{\mathcal M_{\mathrm e4}} &=& \int_0^\infty dr\,u_f(r)u_i(r)\,j_2(Q_fr)\:
\frac{1}{4\pi}\frac{(M_\sigma^2 + \Gamma_\sigma^2/4)^2}
{(X^2-A^2)^2}\:F_2(r) \label{M4}.
\end{eqnarray}
In the above matrix elements, $X$ is defined as $X = \sqrt{M_\sigma^2 + 
\Gamma_\sigma^2/4 - (\omega_a+\omega_b)^2/4}$, while $Y_0(r)$ denotes the 
Yukawa function $e^{-r}/r$. Note that when the value of $k_0^2$ exceeds 
$m_\pi^2$, the analytic continuation $A\rightarrow 
-i\,\sqrt{k_0^2 - m_\pi^2}$~\cite{Chai} is employed for the matrix 
element~(\ref{M2}). Further $u_f(r)$ and $u_i(r)$ denote the reduced 
radial wave functions for the final and initial state heavy quarkonia, 
respectively. The function $F_2(r)$ is defined as
\begin{eqnarray}
F_2(r) &=& \frac{3}{r^3}\left(e^{-Ar} - e^{-Xr}\right) 
+ \frac{3}{r^2}\left(Ae^{-Ar} - Xe^{-Xr}\right) 
+ \frac{1}{r}\left(A^2e^{-Ar} - X^2e^{-Xr}\right) \nonumber \\
&& - \frac{e^{-Xr}(X^2-A^2)}{2}\left(\frac{1+rX}{r}\right),
\end{eqnarray}
a form which is closely related to and in the limit $m_\sigma \rightarrow 
\infty$ actually reduces to a Yukawa $Y_2$ function~\cite{Chai}. It turns 
out that the matrix element~(\ref{M4}) is numerically quite insignificant, 
because of the strong suppression caused by the $j_2$ function for small 
values of $Qr$. Also, the smallness of $k_0$ as compared with 
$\vec k$ precludes the terms proportional to $k_0$ and $k_0^2$ in 
eq.~(\ref{terms}) from playing any major role.

At this point, it is desirable to check the quality of the approximations 
made in obtaining the above matrix elements. This is possible since the 
triple propagator in eq.~(\ref{3prop}) may also be considered without 
any approximation in $k_1$ and $k_2$, at the price of numerically much 
more cumbersome expressions, if one makes use of the Feynman 
parameterization
\begin{equation}
\frac{1}{ABC} = 2\int_0^1 dx\,x \int_0^1 dy 
\frac{1}{[A(1-x)+Bxy+Cx(1-y)]^3}.
\label{feynpar}
\end{equation}
In that case, the amplitude of Fig.~\ref{sigmafig} leads to an expression 
which replaces eq.~(\ref{2q2}), and is of the form
\begin{eqnarray}
T_{\mathrm{2q}} &=& -(8\pi \lambda)^2
\left\{\frac{1}{3}\left(\frac{\vec q\,^2}{4}-Q_f^2\right)
\left[\int_0^1 dx\,x\int_0^1 dy \left\{
{\mathcal M_{\mathrm I}} - A^2{\mathcal M_{\mathrm {II}}}\right\}\right]
\right. \nonumber \\
&& \quad\quad\quad\quad
+ \int_0^1 dx\,x\int_0^1 dy \left(-\frac{\vec q\,^2}{4}(1-2x+xy) - 
Q_f^2(1-xy) + qQ_fz(1-x)\right)
\nonumber \\
&& \quad\quad\quad\quad\quad\quad\quad\quad\quad\quad\quad\:
\left(-\frac{\vec q\,^2}{4}(1-2x+xy) + Q_f^2(1-xy) + qQ_fzx(1-y)\right)
{\mathcal M_{\mathrm {II}}} \nonumber \\ 
&& \quad\quad\quad\:\:\:\:
\left. +\:\omega_a\omega_b\,k_0^2 \int_0^1 dx\,x\int_0^1 dy 
\:{\mathcal M_{\mathrm {II}}}\right\} + T_{\mathrm{exc}},
\label{compl}
\end{eqnarray}
where the matrix elements are given by
\begin{eqnarray}
{\mathcal M_{\mathrm {I}}} &=& 2\int_0^\infty dr\,u_f(r)u_i(r)\,
\frac{(M_\sigma^2 + \Gamma_\sigma^2/4)^2}{8\pi A}\:
e^{-Ar} \nonumber \\
&& j_0\left(r\sqrt{\frac{\vec q\,^2}{4}(1-2x+xy)^2
+Q_f^2x^2y^2 + qQ_fz(1-2x+xy)xy}\:\right), \\
{\mathcal M_{\mathrm {II}}} &=& 2\int_0^\infty dr\,u_f(r)u_i(r)\,
\frac{(M_\sigma^2 + \Gamma_\sigma^2/4)^2}{32\pi A^3}\:
e^{-Ar}(rA+1) \nonumber \\
&& j_0\left(r\sqrt{\frac{\vec q\,^2}{4}(1-2x+xy)^2
+Q_f^2x^2y^2 + qQ_fz(1-2x+xy)xy}\:\right).
\end{eqnarray}
Here the term proportional to $k_0^2$ is again only of minor importance.
Note that in order to obtain the contribution $T_{\mathrm{exc}}$ to 
eq.~(\ref{compl}), it is necessary to make the substitution $k_a 
\leftrightarrow k_b$, which implies $\omega_a \leftrightarrow \omega_b$ 
and $Q_f \rightarrow -Q_f$. In the above matrix elements, the quantity $A$ 
is now defined as $A = \sqrt{m_*^2 - K_0^2}$, involving an effective mass 
$m_*$ and an energy transfer variable $K_0$. These are defined according 
to
\begin{eqnarray}
m_*^2 &=& \left(M_\sigma^2 + \frac{\Gamma_\sigma^2}{4}\right)(1-xy) 
- m_\pi^2 x(2(1-x)(1-y) -xy^2) + 2\left(\frac{\vec q\,^2}{4} - Q_f^2\right) 
x(1-x)(1-y), \\
K_0 &=& \omega_a(1-x) -\omega_b x(1-y) + k_0,
\end{eqnarray}
where $M_\sigma$ again denotes the pole position 
$m_\sigma-i\Gamma_\sigma/2$ of the 
sigma resonance, and $k_0$ is taken to be $(\omega_b - \omega_a)/2$.

All the matrix elements for the pion exchange 
amplitudes have here been considered in the nonrelativistic limit, even 
though it was noted that that limit is not realistic in case of the single 
quark amplitudes. The employment of the nonrelativistic limit is expected 
to be permissible here since the Yukawa functions which arise from the 
propagators of the exchanged pions and $\sigma$ mesons negate the delicate 
orthogonality that is inherent in the quarkonium wave functions. Thus 
relativistic effects constitute only a 
correction to the pion-exchange matrix elements, and are expected to be 
rather small because of the relatively large constituent masses of the 
charm and bottom quarks. If, however, it is desirable to consider 
relativistic effects for the pion-exchange amplitudes, a relativistic 
matrix element analogous to eq.~(\ref{relmat}) may be constructed 
according to
\begin{equation}
{\mathcal M_{\mathrm{exch}}^{\mathrm{rel}}} = \int 
\frac{d^3P}{(2\pi)^3}\frac{d^3k}{(2\pi)^3}\:
\varphi_f^*\left(\vec P + \frac{\vec Q}{2} - \frac{\vec k}{2}\right)\:
T_{\mathrm{2q}}(\vec k,\vec P)\:
\varphi_i \left(\vec P - \frac{\vec Q}{2} + \frac{\vec k}{2}\right),
\label{relexchmat}
\end{equation}
where the amplitude $T_{\mathrm{2q}}(\vec k,\vec P)$ corresponds to 
eq.~(\ref{2q}) multiplied by the $\sigma$ factors from 
eq.~(\ref{sigmares}) and subject to the approximations that led to 
eq.~(\ref{2q3}). Furthermore, $T_{\mathrm{2q}}(\vec k,\vec P)$ should also 
contain the quark and antiquark bispinors analogous to eq.~(\ref{alpha}).
In the matrix element~(\ref{relexchmat}), the wavefunctions are defined 
according to
\begin{eqnarray}
\varphi_f^* &=& \int d^3r'\: 
e^{i\left(\vec P + \frac{\vec Q}{2} - \frac{\vec k}{2}\right)
\cdot\vec r\,'} \varphi_f^*(\vec r\,'\,), \\
\varphi_i &=& \int d^3r \: 
e^{-i\left(\vec P - \frac{\vec Q}{2} + \frac{\vec k}{2}\right)
\cdot\vec r} \varphi_i(\vec r\,).
\end{eqnarray}

Numerical evaluation and comparison of the approximate model for the pion 
exchange diagrams with the version based on the Feynman 
parameterization~(\ref{feynpar}) indicates that the approximate model 
successfully describes the pion exchange amplitudes, with only very small 
deviations from eq.~(\ref{compl}). To summarize, a very good description 
of the pion exchange diagrams may already be obtained by considering only 
the first term in eq.~(\ref{2q3}). The results presented in the next 
section correspond, unless otherwise indicated, to eq.~(\ref{1q2}) for the 
single quark amplitudes, and to eq.~(\ref{2q3}) with exception of the term 
proportional to $k_0^2$ for the pion exchange amplitudes.

\newpage

\section{Results for $\Upsilon' \rightarrow \Upsilon\,\pi\pi$ and $\psi' 
\rightarrow J/\psi\,\pi\pi$}

\begin{figure}[h!]
\begin{center}
\epsfig{file = 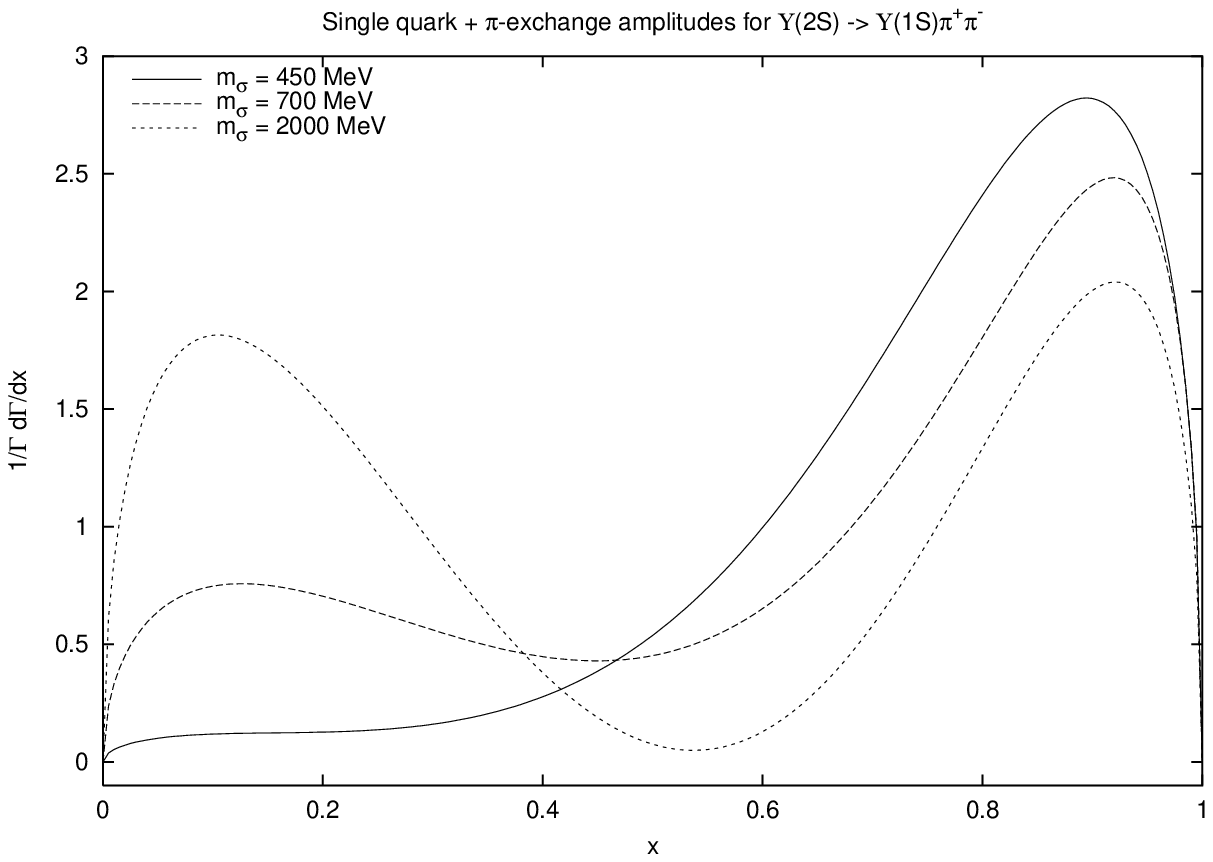}
\epsfig{file = 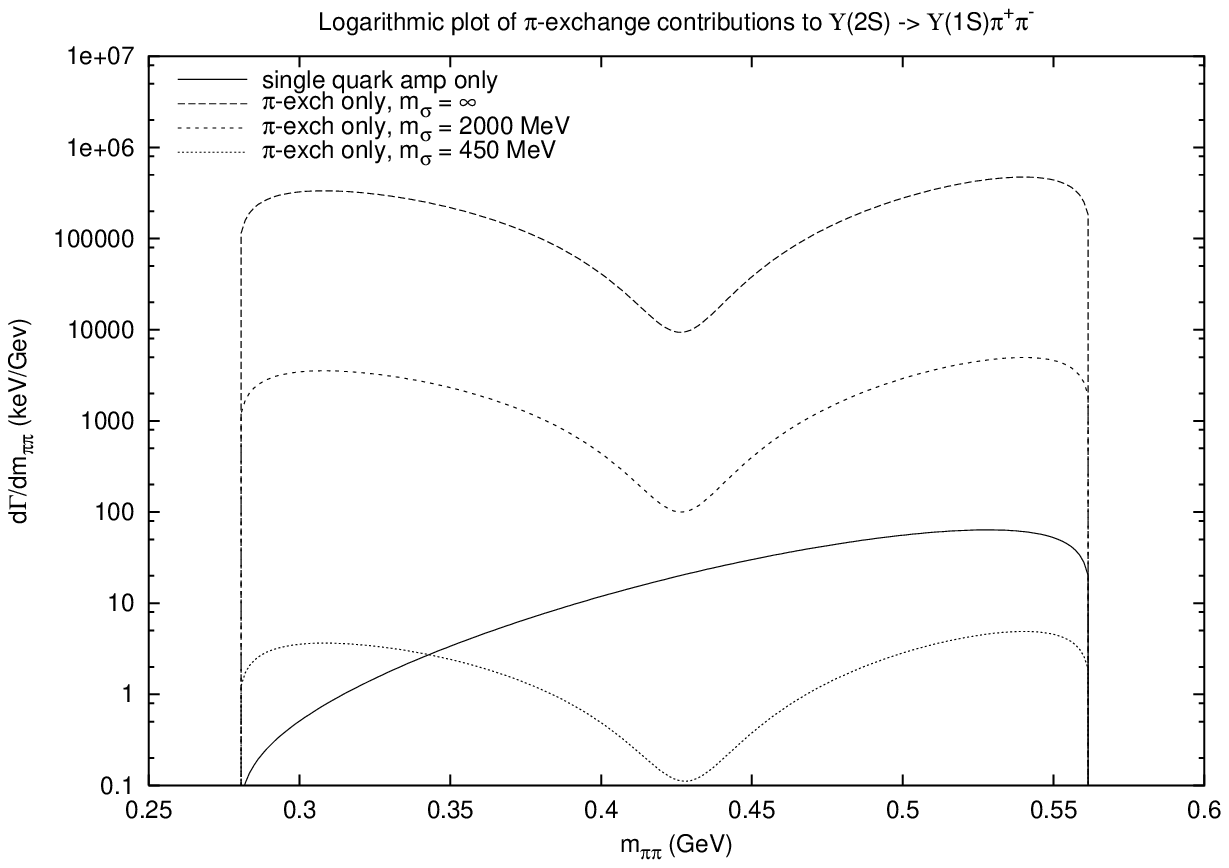}
\caption{Illustration of the reduction of $\pi$-exchange amplitudes 
brought about by the introduction of an intermediate $\sigma$ meson with a 
width of $\Gamma_\sigma$ = 550 MeV for $\lambda = +0.02 \:\mathrm{fm}^3$. 
Note that the relativistic single quark amplitude~(\ref{relmat}) begins to 
dominate when $m_\sigma$ is less than $\sim 1$ GeV.}
\label{exchfig}
\end{center}
\end{figure}

The pion exchange mechanisms in eq.~(\ref{ampl}) may $\it a~priori$ give 
large contributions to the decays $\Upsilon' \rightarrow \Upsilon\,\pi\pi$ 
and $\psi'\rightarrow J/\psi\,\pi\pi$ because they are not suppressed by 
the orthogonality of the $2S$ and $1S$ wave functions of the heavy 
quarkonium states. Therefore, the contribution from the pion exchange 
diagrams is expected to be dominant if point couplings to heavy 
quarks are employed. This is illustrated in Fig.~\ref{exchfig}, where the 
case of 
$m_\sigma = \infty$ corresponds to pointlike couplings. Also, if the 
$\pi\pi$ decay of heavy quarkonium states is dominated by pion exchange 
mechanisms, the widths of the bottomonium states are expected to be larger 
than those of the corresponding charmonium states, which is in 
conflict with experiment~\cite{PDG,bbexp}. This is the case because 
of the narrowness of the $b\bar b$ wavefunctions combined with the short 
range nature of the Yukawa functions in the pion exchange amplitude.
This suggests that the $Q\pi\pi$ coupling is mediated by a light scalar 
meson or glueball. 

Indeed, if a $\sigma$ meson lighter than 1 GeV is 
employed in combination with a relativistic treatment of the single quark 
amplitudes in eq.~(\ref{ampl}), then the effects of pion exchange diagrams 
may be reduced, and their contributions may in fact become subdominant as 
compared to the single quark amplitudes, which allows agreement with 
experiment. The results presented in this section indicate that $\sigma$ 
masses of the order $\sim 500$ MeV lead to a favorable description of the 
current experimental data on the $\pi\pi$ decays of the $2S$ states of 
heavy quarkonia.
 
The calculated widths and two-pion energy distributions were obtained by 
simultaneously optimizing the results for $\Upsilon' \rightarrow 
\Upsilon\,\pi^+\pi^-$ and $\psi' \rightarrow J/\psi\,\pi^+\pi^-$. 
Furthermore, even though the coupling constant $\lambda$ is in principle a 
free parameter, it is strongly constrained by the shape of the $\pi\pi$ 
energy distribution. This reflects the fact that the pion-exchange 
amplitudes may contribute significantly alongside the single quark 
diagrams. It is therefore necessary to choose $\lambda$ in such 
a way that not only the width but also the $\pi\pi$ energy spectrum is 
optimally described.

\begin{figure}[h!]
\begin{center}
\epsfig{file = 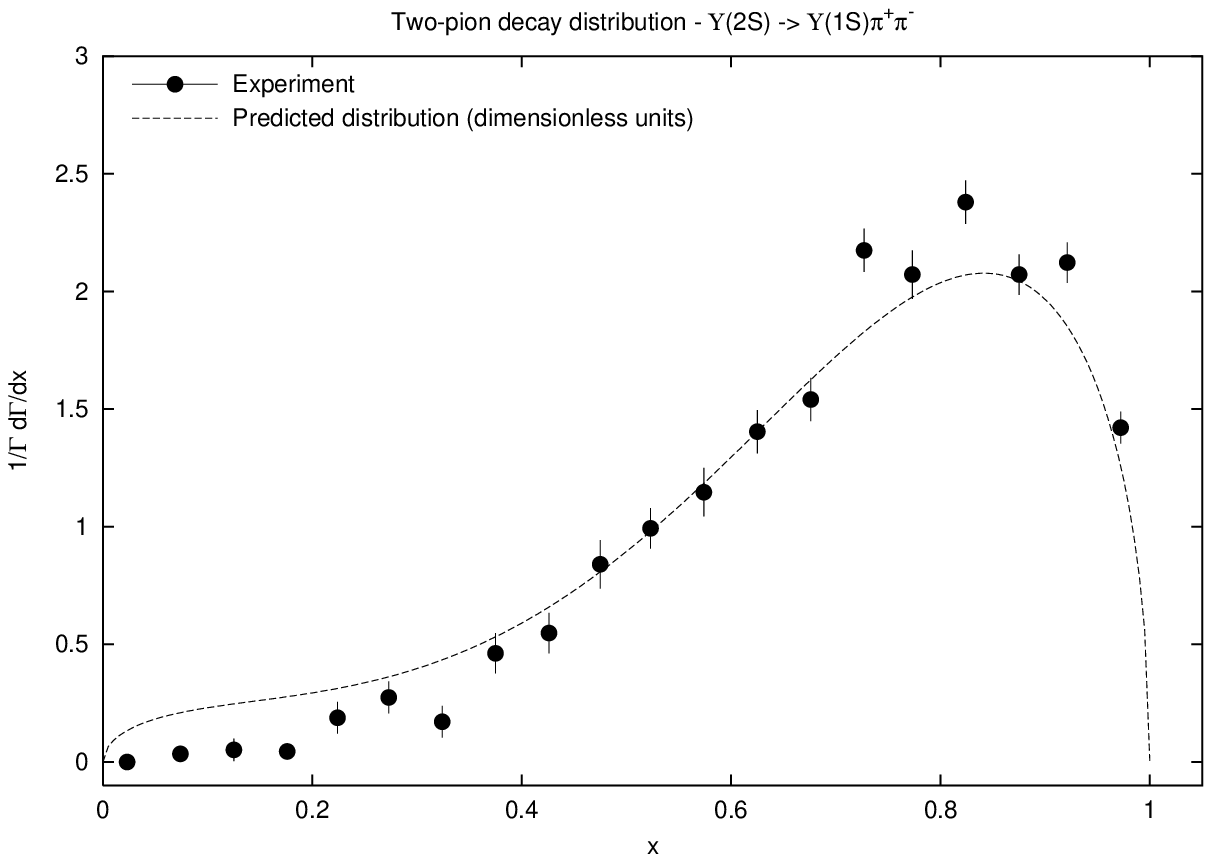}
\epsfig{file = 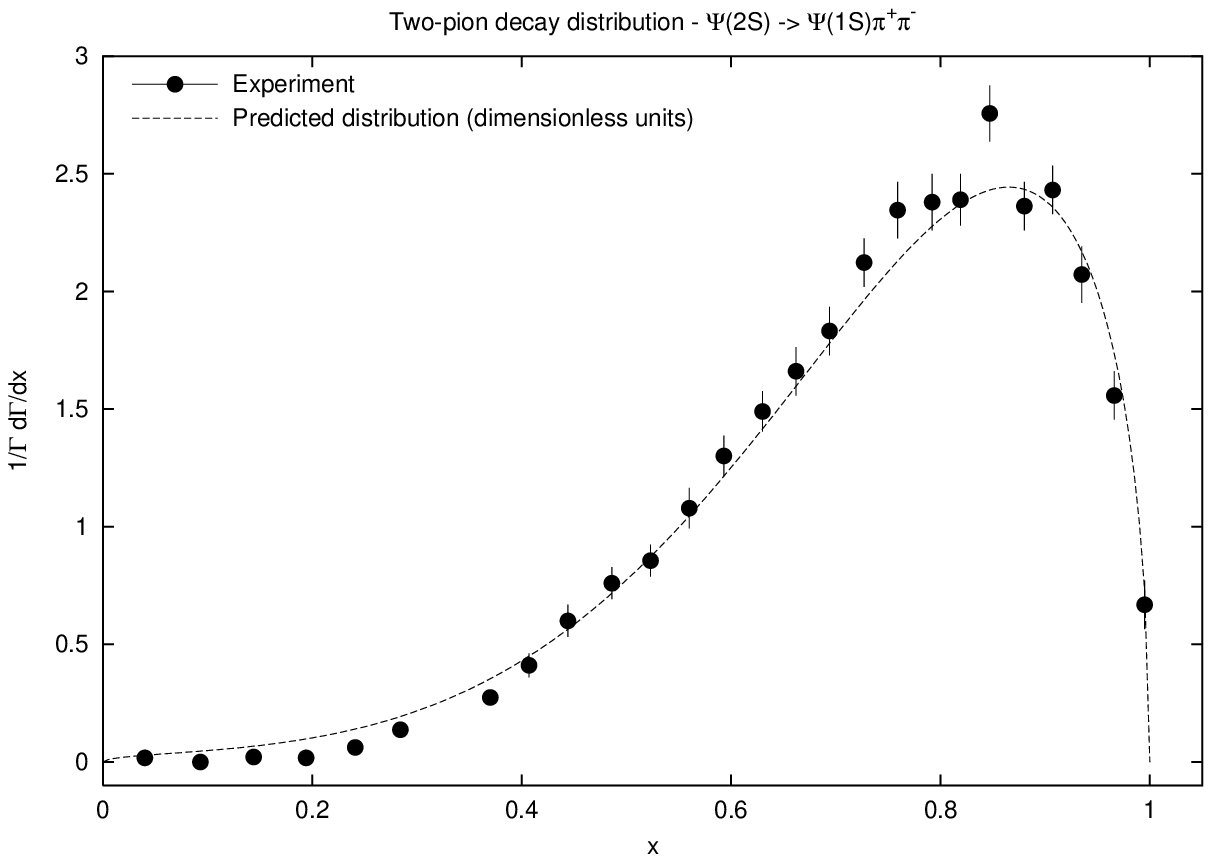}
\caption{Comparison of calculated and experimental~\cite{bbexp} $\pi\pi$ 
energy distributions for
$\Upsilon' \rightarrow \Upsilon\pi^+\pi^-$ and $\psi' \rightarrow
J/\psi\,\pi^+\pi^-$, for $m_\sigma$ = 450 MeV, $\Gamma_\sigma$ = 550 MeV 
and \mbox{$\lambda = -0.02\:\mathrm{fm}^3$.} The calculated width for 
$\pi^+\pi^-$ decay is 5.89 keV for $b\bar b$ and 53.5 keV for $c\bar 
c$. The scaled $\pi\pi$ invariant mass $x$ is defined in 
eq.~(\ref{xvar}).}
\label{expdat}
\end{center}
\end{figure}

\begin{figure}[h!]
\begin{center}
\epsfig{file = 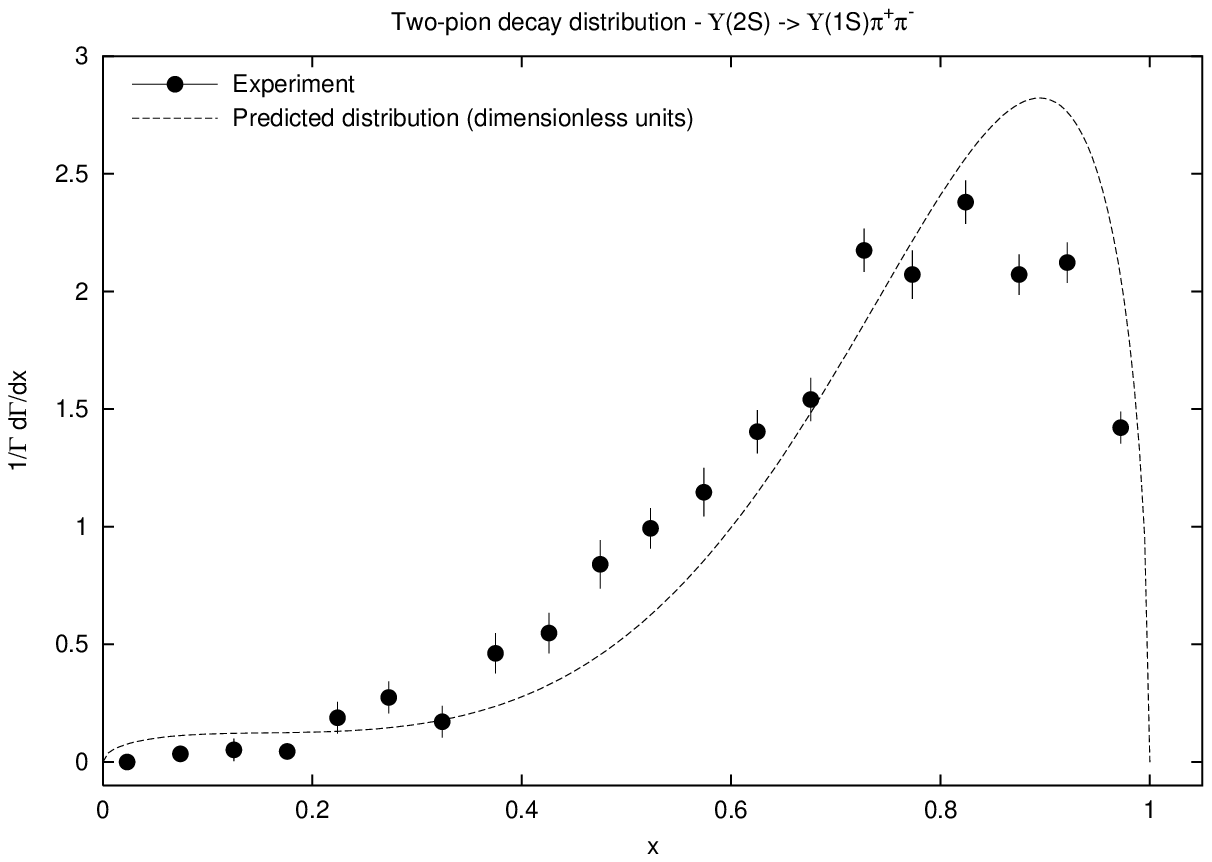}
\epsfig{file = 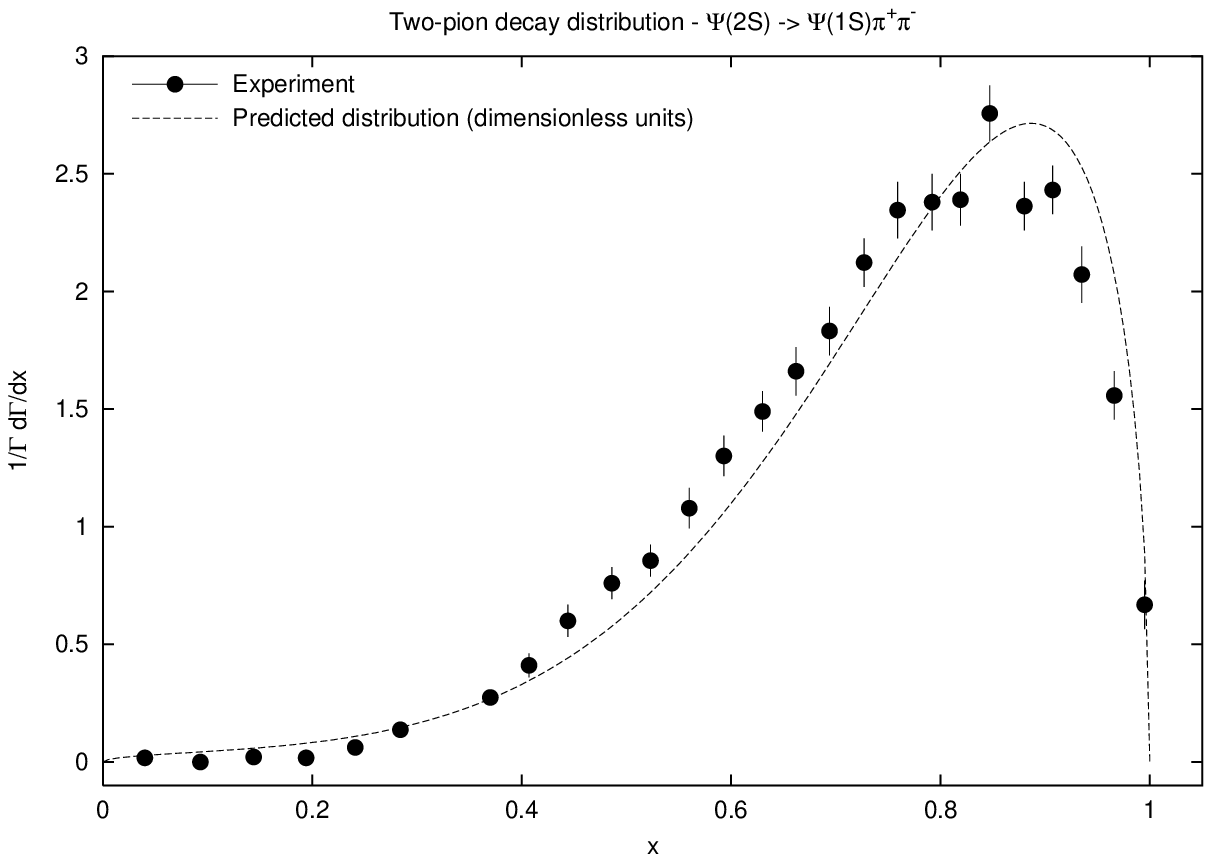}
\caption{Comparison of calculated and experimental~\cite{bbexp} $\pi\pi$ 
energy distributions for $\Upsilon' \rightarrow \Upsilon\pi^+\pi^-$ and 
$\psi' \rightarrow J/\psi\,\pi^+\pi^-$, for $m_\sigma$ = 450 MeV, 
$\Gamma_\sigma$ = 550 MeV and \mbox{$\lambda = +0.02\:\mathrm{fm}^3$.} 
The calculated width for $\pi^+\pi^-$ decay is 9.54 keV for $b\bar 
b$ and 67.5 keV for $c\bar c$. The scaled $\pi\pi$ invariant mass $x$ is 
defined in eq.~(\ref{xvar}).}
\label{expdat2}
\end{center}
\end{figure}

The results, which yielded the $\sigma$ meson parameters $m_\sigma$ = 450 MeV 
and $\Gamma_\sigma$ = 550 MeV, are shown in Figs.~\ref{expdat} 
and~\ref{expdat2} for negative and positive values of $\lambda$, 
respectively. The calculated decay widths corresponding to $\lambda = 
-0.02$ are shown Table~\ref{restab}. This negative value of $\lambda$ was 
found to lead to an optimal description of both the $b\bar b$ and $c\bar 
c$ data. It is noteworthy that the sensitivity to the sign of $\lambda$ is 
entirely due to the presence of pion exchange amplitudes in 
eq.~(\ref{ampl}). The masses of the heavy quarkonium states correspond to 
those given by ref.~\cite{PDG}. Using instead the energy eigenvalues of 
Table~\ref{spektrtab} corresponding to the wave functions obtained by 
solving the BSLT equation makes very little difference since they agree 
quite well with the experimental masses. The pion masses were taken to be 
139.57 MeV for the charged pions, and 134.98 MeV for the neutral pion. The 
heavy constituent quark masses used were those listed in 
Table~\ref{partab}. Unless otherwise indicated, the results in this 
section refer to $\pi^+\pi^-$ decays. 

\vspace{0.35cm}

\begin{table}[h!]
\begin{center}
\begin{tabular}{c||c|c|c|c|c} 
Decay & $\Gamma_{\mathrm {tot}}$ & Br.Rat. & $\Gamma_{\mathrm {exp}}$
& $\Gamma_{\mathrm {calc}}$ & $q_{\mathrm {max}}$ \\ \hline\hline
& & & & & \\
$\Upsilon'\rightarrow \Upsilon\:\pi^+\pi^-$ & 44 $\pm$ 7 keV & 
18.8 $\pm$ 0.6 \% & 8.3 $\pm$ 1.3 keV & 5.89 keV & 475 MeV \\
$\Upsilon'\rightarrow \Upsilon\:\pi^0\pi^0$ &                & 
9.0 $\pm$ 0.8 \%  & 4.0 $\pm$ 0.8 keV & 3.07 keV & 480 MeV \\ 
& & & & & \\
$\psi'\rightarrow J/\psi\:\pi^+\pi^-$ & 277 $\pm$ 31 keV     & 
31.0 $\pm$ 2.8 \% & 86 $\pm$ 12 keV   & 53.5 keV & 477 MeV \\
$\psi'\rightarrow J/\psi\:\pi^0\pi^0$ &                      & 
18.2 $\pm$ 2.3 \% & 50 $\pm$ 10 keV   & 27.8 keV & 481 MeV \\
\end{tabular}
\caption{Experimental data and calculated widths for $\pi\pi$ decays of 
the $\Upsilon'$ and $\psi'$ mesons, for $\lambda = -0.02\:\mathrm{fm}^3$, 
$m_\sigma$ = 450 MeV and $\Gamma_\sigma$ = 550 MeV. Experimental total 
widths, branching fractions and resulting widths for $\pi^+\pi^-$ and 
$\pi^0\pi^0$ are given. The column $q_{\mathrm {max}}$ lists the maximal 
momenta $|\vec q\,|$ attainable by the pion pairs.} \label{restab}
\end{center}
\end{table}

\newpage

\newpage

\newpage

\begin{figure}[h!]
\begin{center}
\epsfig{file = 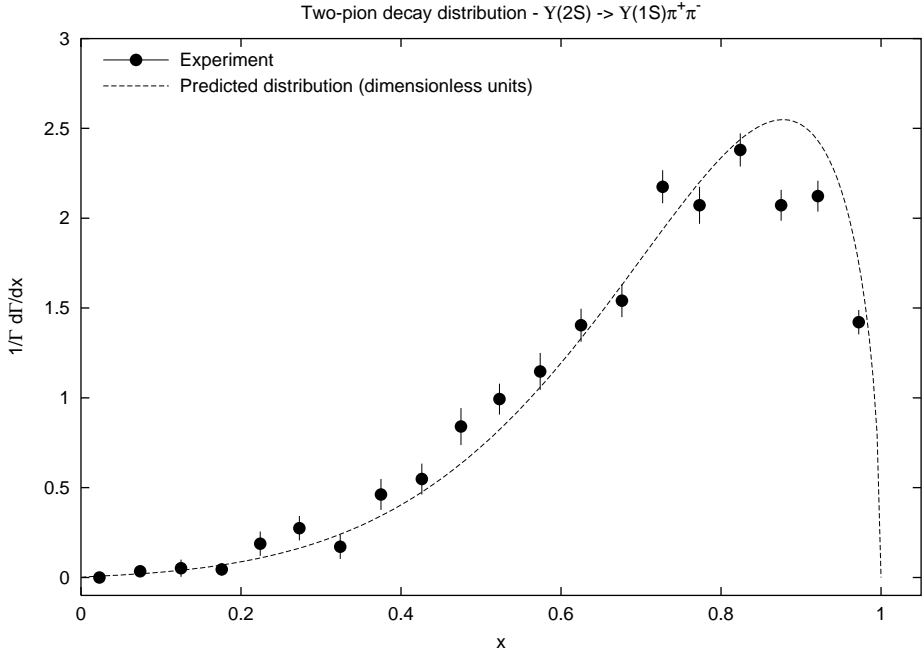}
\caption{Comparison of calculated and experimental~\cite{bbexp} $\pi\pi$ 
energy distributions for $\Upsilon' \rightarrow \Upsilon\pi^+\pi^-$, with 
$m_\sigma$ = 450 MeV, $\Gamma_\sigma$ = 550 MeV and \mbox{$|\lambda| = 
0.02\:\mathrm{fm}^3$.} The calculated $\pi^+\pi^-$ width is 7.09 keV. The 
scaled $\pi\pi$ invariant mass $x$ is defined in eq.~(\ref{xvar}).
This result is obtained when only the single quark diagrams are retained 
in the calculations.}
\label{singleq}
\end{center}
\end{figure}

The results presented in Figs.~\ref{expdat} and~\ref{expdat2} indicate 
that a more favorable description of the $\pi\pi$ invariant mass 
distributions may be obtained if the coupling constant $\lambda$ is taken 
to be negative. This appears to be in line with the fact that the 
experimental distributions for $b\bar b$ and $c\bar c$ as shown in 
Figs.~\ref{expdat} and~\ref{expdat2} show slight differences. In 
particular, the peak at high $x$ is somewhat lower for $b\bar b$, while 
the tail at low $x$ is broader for $b\bar b$. It has also been noted in 
ref.~\cite{bbexp} that the resonance model of ref.~\cite{Schwinger} cannot 
be simultaneously fitted to both the $b\bar b$ and $c\bar c$ data. It is 
seen from Fig.~\ref{expdat} that if negative values of $\lambda$ are 
employed then the abovementioned qualitative differences between the 
$b\bar b$ and $c\bar c$ decay distributions may be accounted for. The 
obtained value for the $\sigma$ meson mass, 450 MeV, is constrained even 
if single quark diagrams only are considered. In that case, a higher $\sigma$ 
mass would lead to a distribution which is peaked too far to the right. If 
pion exchange amplitudes are considered as well, a higher $\sigma$ mass 
leads to unrealistically large pion exchange contributions. The results 
are not very sensitive to the $\sigma$ width, but standard values of $\sim 
500$ MeV appear to be favored by the calculations.

It is also evident from Fig.~\ref{expdat} that the pion exchange 
contribution, while having an overall favorable effect, appears to be 
somewhat overpredicted, particularly for $b\bar b$. This problem can be 
traced to the nonrelativistic treatment of the pion exchange contribution, 
and can be alleviated if relativistic matrix elements are employed 
according to eq.~(\ref{relexchmat}). This relativistic weakening of the 
pion exchange amplitudes may also remedy the apparent underprediction of 
the $\Upsilon'\rightarrow \Upsilon\:\pi^+\pi^-$ width as given by 
Table~\ref{restab}. For $c\bar c$ the results are much less sensitive to 
the exact strength and form of the pion exchange amplitude, an effect 
which is due to the much broader radial wavefunctions of the $c\bar c$ 
states. In case of the decay $\psi'\rightarrow J/\psi\:\pi^+\pi^-$, it may 
actually be desirable to employ a $20-30\,\%$ larger value of $\lambda$, 
as indicated e.g. in ref.~\cite{Mannel}.

\newpage

\newpage

As expected, the logarithmic plot of the pion exchange amplitudes in 
Fig.~\ref{exchfig} reveals that if point couplings between heavy 
constituent quarks and two-pions are employed, then pion exchange will be 
the dominant mechanism for $\pi\pi$ decay of the heavy quarkonium states. 
Fig.~\ref{exchfig} also shows that for the values of $m_\sigma$ obtained 
by fitting the model to experimental data, the pion exchange amplitudes 
are subdominant and constitute a relatively small correction. However, the 
contribution from pion exchange may still not be neglected. By comparing 
Figs.~\ref{expdat} and~\ref{expdat2}, it is revealed that negative values 
of $\lambda$ allow for better agreement with experiment.

\begin{figure}[h!]
\begin{center}
\epsfig{file = 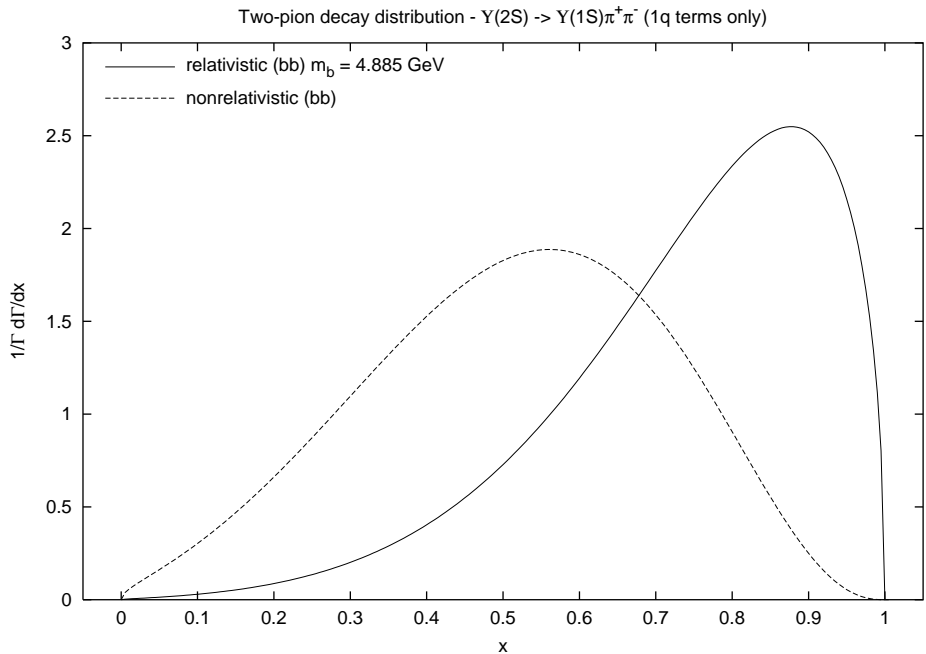}
\caption{Comparison between relativistic and nonrelativistic results for 
bottomonium ($b\bar b$), when only the single quark amplitudes are 
considered. The results correspond to a $\sigma$ mass of 450 MeV and width 
of 550 MeV, and a coupling constant of $|\lambda|$ = 0.02. The
scaled $\pi\pi$ invariant mass $x$ is defined in eq.~(\ref{xvar}). The 
relativistic and nonrelativistic forms give $\pi^+\pi^-$ widths of 7.09 
keV and 0.47 keV, respectively. In case of charmonium ($c\bar c$), the 
nonrelativistic distribution is qualitatively similar, but somewhat 
shifted to the right.}
\label{1qfig}
\end{center}
\end{figure}

Fig.~\ref{1qfig} illustrates the fact that, even though the charm 
and bottom constituent quarks are heavy, the nonrelativistic limit is not 
applicable for the single quark diagrams. This comes about because of the 
orthogonality of the wavefunctions in the matrix element~(\ref{nrmat}). 
It follows that the single quark mechanisms are highly suppressed 
since the product $qr$ is always small for the observables
considered here. If the spinor factors of eq.~(\ref{alpha}) in the 
relativistic matrix element~(\ref{relmat}) are taken into account, they 
will lead to an increase of the contributions from single quark diagrams.
The resulting amplitude then becomes almost constant over the whole range 
of two-pion momenta covered by the $\Upsilon'\rightarrow \Upsilon\pi\pi$ 
decays, allowing agreement with the phenomenological deductions of earlier 
work~\cite{Schwinger,Mannel}. Also, since the relativistic modifications 
increase the effect of the single quark amplitudes by a factor $\sim 10$, 
they also allow avoidance of large values of the coupling 
constant $\lambda$. This is even more important since otherwise, even 
though the pion exchange contributions are very much suppressed by 
inclusion of the $\sigma$ meson, they may still overwhelm the single quark 
contributions in the nonrelativistic limit, making it very difficult to 
obtain agreement with the experimental $\pi\pi$ energy distributions. It 
is worth noting in this context that the nonrelativistic approximation 
becomes realistic only if the constituent quark mass is larger than $\sim 
30$ GeV.

\begin{figure}[h!]
\begin{center}
\epsfig{file = 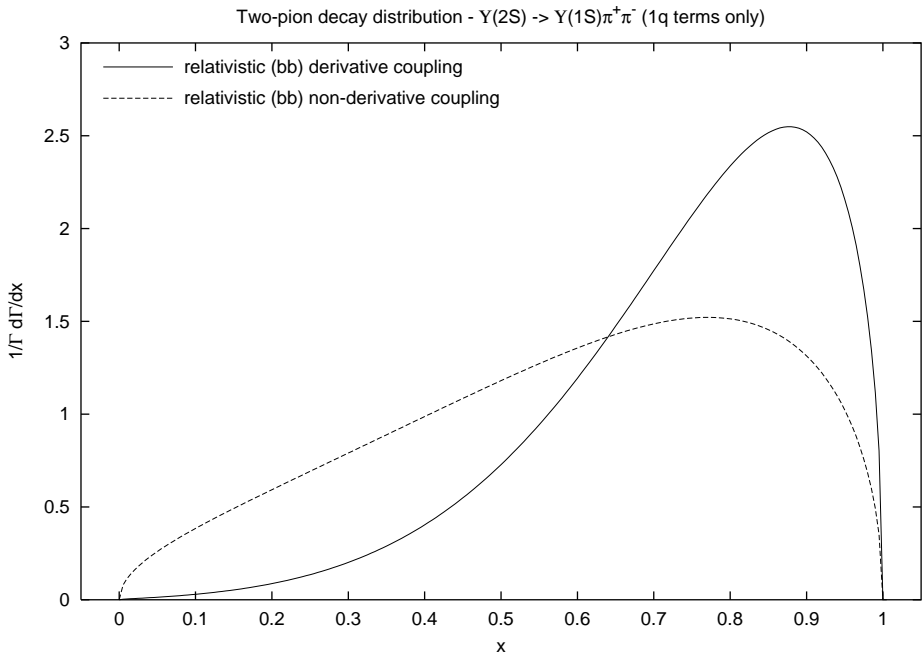}
\caption{Comparison between derivative and non-derivative couplings in 
the relativistic case (single quark diagrams only) for $b\bar b$. The
scaled $\pi\pi$ invariant mass $x$ is defined in eq.~(\ref{xvar}). The 
results correspond to the same set of parameters as Fig.~\ref{1qfig}.} 
\label{noder}
\end{center}
\end{figure}

Another interesting point is to check whether the derivative coupling to 
pions in eq.~(\ref{Q2pi}) actually does provide a superior description of 
the $\pi\pi$ energy spectrum. A comparison between a model with derivative 
couplings and one without them is given for the single quark amplitudes in 
Fig.~\ref{noder}. As the inclusion of the $\sigma$ mesons leads to the 
same conclusions of the vanishing of the pion exchange contributions for 
both models, a fair comparison may already be obtained by comparing the 
single quark amplitudes only. Fig.~\ref{noder} indicates that even in the 
relativistic case, a model without derivative couplings on the pion fields 
does not provide a good description of the $\pi\pi$ distribution. Such a 
model is seen to give a rather structureless distribution, which does not 
reproduce the sharp peak at large values of the scaled $\pi\pi$ invariant 
mass $x$ and also gives a wrong curvature at low $x$, a result which has 
also been obtained by ref.~\cite{Schwinger}. This is reassuring since the 
derivative couplings for pions are consistent with the role of pions as 
Goldstone bosons of the spontaneously broken approximate chiral symmetry 
of QCD.

\newpage

\section{Discussion}

The results of the previous section indicate that an overall satisfactory 
description of the decays $\Upsilon' \rightarrow \Upsilon\,\pi\pi$ and 
$\psi' \rightarrow J/\psi\,\pi\pi$ has been achieved. This good agreement 
was shown to depend on several 
factors, most notably the use of derivative couplings to pions, 
relativistic treatment of amplitudes where the two-pions are emitted from 
a single constituent quark, and the inclusion of an intermediate 
scalar $\sigma$ meson with a mass of $\sim 500$ MeV. Inclusion of the 
$\sigma$ meson has been shown to lead to a drastic reduction of the
amplitudes, where one of the emitted pions is rescattered by the other 
constituent quark. Still, there remains some issues with these 
transitions that deserve further discussion. 

As seen from Table~\ref{restab}, the decay widths for the $c\bar c$ system are 
generally underpredicted by $\sim 30-40$~\%, if the coupling constant 
$\lambda$ is determined from the corresponding decays in the $b\bar b$ 
system. There appears to be no obvious reason for this 
underprediction, as the results are not exceedingly sensitive to the 
particulars of the model used, nor to the exact values of the masses of 
the constituent quarks involved. Furthermore, the pion exchange amplitudes 
are relatively insignificant for the $c\bar c$ system, and cannot account 
for this underprediction. If the value of $\lambda$ is increased by $\sim 
20$~\% for the $c\bar c$ system, the computed width for the 
transition $\psi'\rightarrow J/\psi\,\pi^+\pi^-$ comes into close 
agreement with the current experimental data. A similar requirement 
for a slight increase of the coupling constant for the $c\bar c$ system 
has also been noted by 
ref.~\cite{Mannel}, and is therefore likely to be a real effect. 
Nevertheless, it has to be noted that the total width of the 
$\psi'$~\cite{PDG} state is difficult to determine experimentally and 
still remains somewhat uncertain. 

It is also noteworthy that the agreement between the calculated and 
experimental $\pi\pi$ energy distributions, as shown in Fig.~\ref{expdat}, 
turns out to be somewhat better for $c\bar c$ than for $b\bar b$. A likely 
explanation for this feature is the nonrelativistic treatment of the pion 
exchange amplitudes, eq.~(\ref{2q3}). This~\mbox{results} in a slight 
overprediction of the pion exchange contribution, the effect of which is 
amplified by the narrowness of the $b\bar b$ wavefunctions. As the present 
model has only few degrees of freedom, and 
does not really constitute a "fit" to the available data (e.g. the quark 
masses are fixed by the $Q\bar Q$ spectrum) then lack of perfect agreement 
with experiment is not unexpected. A somewhat better fit to the available 
data on $\pi\pi$ decay in the $c\bar c$ and $b\bar b$ systems has been 
obtained in ref.~\cite{Mannel} by employment of a more complex Lagrangian 
with more adjustable coupling constants. Another possible explanation is 
that eq.~(\ref{sigmares}) may only be a very crude model for the 
interacting $\pi\pi$ state. Indeed, ref.~\cite{Schwinger} has employed a 
somewhat more sophisticated $\sigma$ model where the width of the 
resonance is not a constant. A further possibility is that two-quark 
contributions associated with intermediate negative energy quarks~(see 
e.g. the "Z" diagrams of ref.~\cite{tl2pi}) may significantly modify the 
calculated decay distributions and widths. These effects, which depend 
explicitly on the form and Lorentz coupling structure of the 
quark-antiquark interaction arise from the elimination of negative energy 
components in the BSLT quasipotential reduction, which has been employed 
in this work. For e.g. the scalar confining interaction and the 
coupling~(\ref{Q2pi}) they can be shown to be proportional to $M_Q^{-3}$ 
at the nonrelativistic limit and are therefore highly relativistic effects 
that, when treated properly, are expected to be of minor significance.

An outstanding problem with the hadronic decays of heavy quarkonia 
has been the theoretical understanding of the $\Upsilon(3S) 
\rightarrow \Upsilon\,\pi\pi$ and $\Upsilon(3S) \rightarrow 
\Upsilon'\,\pi\pi$ decays. These have been experimentally studied in 
ref.~\cite{3Sexp}, and were found to have roughly equal widths of $\sim 1$ 
keV. This is a problematic feature since the derivative coupling of the 
pions in combination with the phase space factors naturally lead to a 
strong suppression of the latter decay mode relative to the former. This 
has led to the introduction of different coupling constants for the 
different decay modes in order to compensate for the increase or decrease 
in phase space. While this may be reasonable if the $\Upsilon$ and $\psi$ 
states are treated as fundamental particles, it is difficult to formulate a 
consistent model using variable coupling constants if the $\pi\pi$ decay 
is modeled in terms of a $Q\pi\pi$ coupling. Despite the generally good 
agreement with experiment obtained for the $\pi\pi$ decays of the 
$\Upsilon'$ and $\psi'$ states, a width of only 1 keV for 
$\Upsilon(3S)\rightarrow \Upsilon\,\pi\pi$ cannot be achieved with the 
present set of parameters.

\newpage

\begin{figure}[h!]
\begin{center}
\epsfig{file = 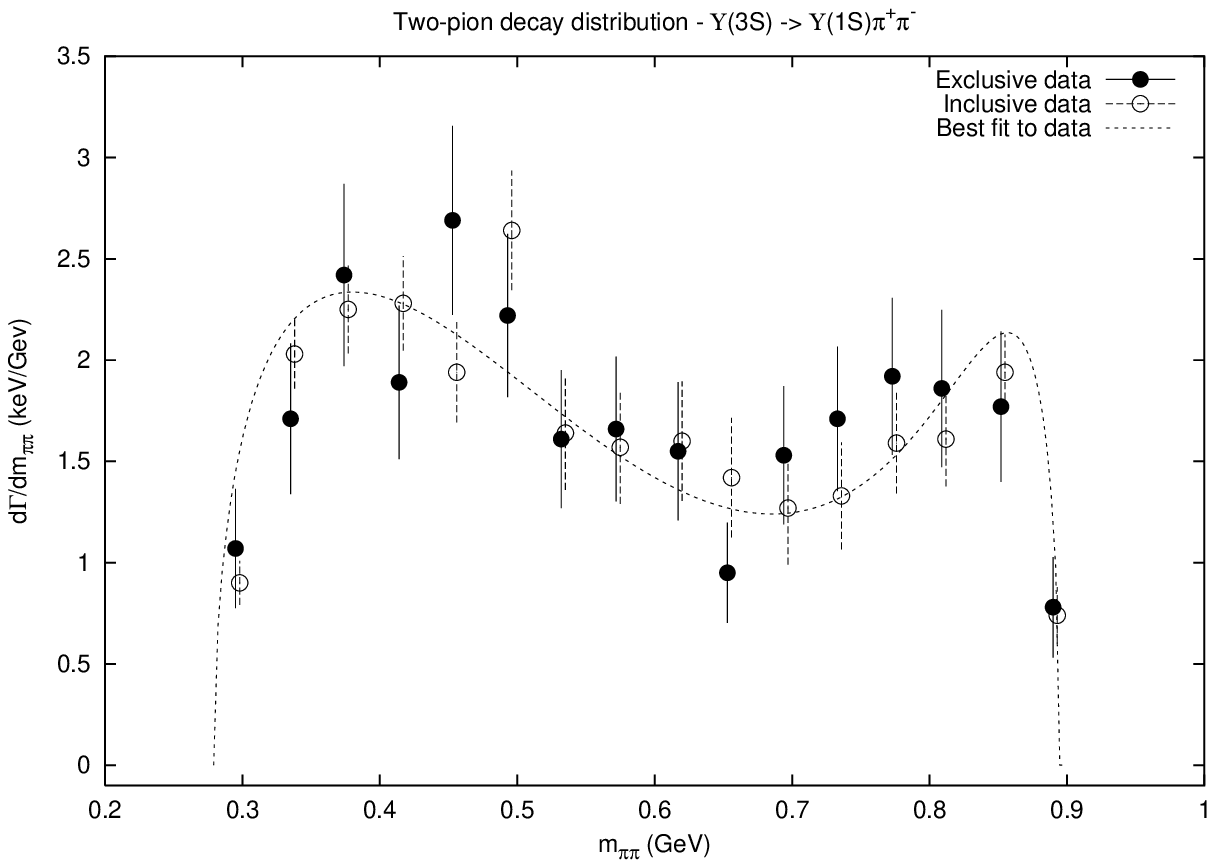}
\epsfig{file = 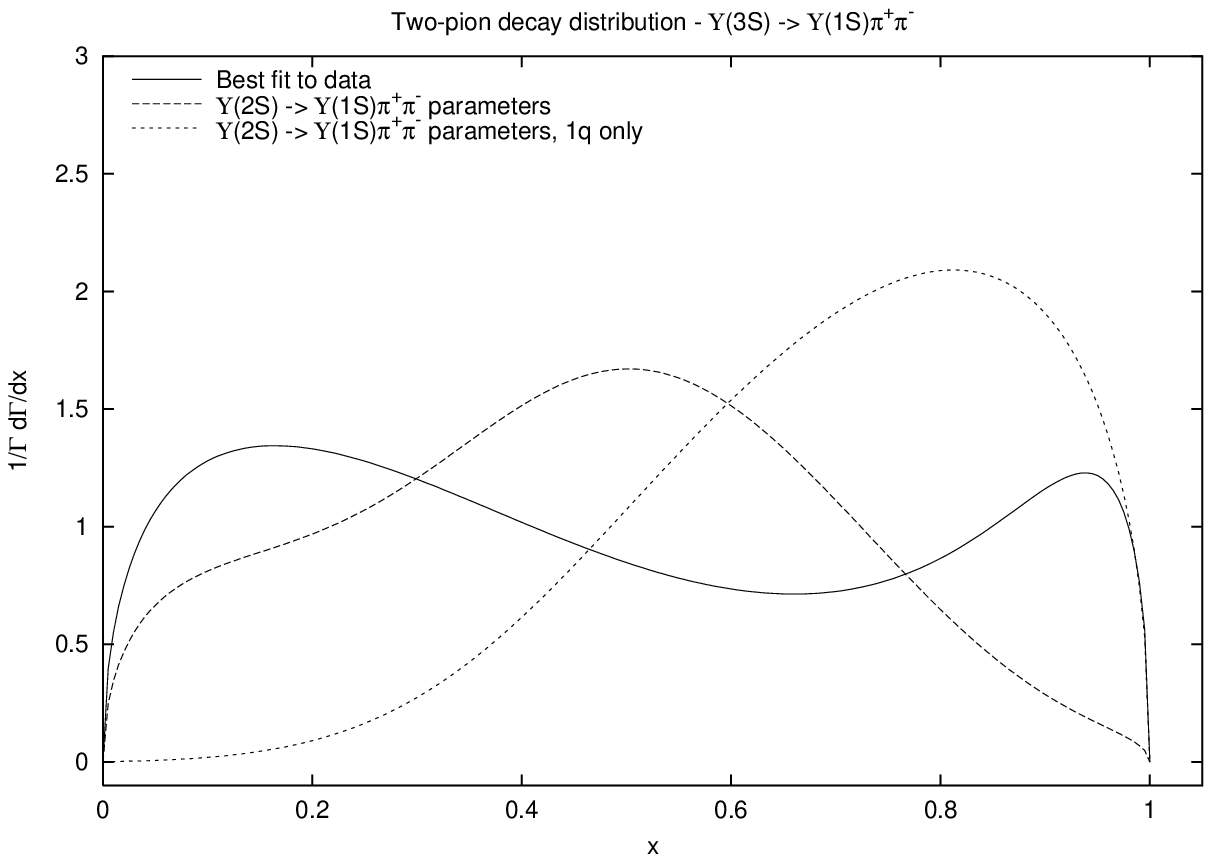}
\caption{Best fit to data~\cite{3Sexp} for the decay 
$\Upsilon(3S)\rightarrow \Upsilon\,\pi^+\pi^-$ and comparison with results 
using the parameters from Table~\ref{restab}. The parameters obtained were 
$m_\sigma$ = 
1400~MeV, $\Gamma_\sigma$ = 100 MeV, $\lambda = 2.7\cdot 10^{-3}$, and 
give $\Gamma_{\pi^+\pi^-}$ = 1.07 keV. The scaled $\pi\pi$ invariant mass 
$x$ is defined in eq.~(\ref{xvar}).}
\label{3Sfig}
\end{center}
\end{figure}

\newpage

The $\pi\pi$ energy distribution of the $\Upsilon(3S) 
\rightarrow \Upsilon\,\pi\pi$ decay shows an anomalous double-peaked 
structure, which cannot be explained by models dominated by single-quark 
amplitudes, such as the one employed above for the $\pi\pi$ decays of the
$\Upsilon'$ and $\psi'$ states. This is in obvious contradiction with 
the conclusions from analysis of the $\pi\pi$ energy distributions of the 
$\Upsilon'\rightarrow \Upsilon\,\pi\pi$ decays. In this situation it 
seems natural to assume that the initial $\Upsilon(3S)$ state may
have a more complex structure than a simple radial excitation of the 
$\Upsilon(1S)$. This has been investigated in 
refs.~\cite{BB1,BB2}, where acceptable agreement with the experimentally 
observed $\pi\pi$ energy distributions has been attained by consideration 
of intermediate $B\bar B^*$ states. However, the explicit calculation of 
coupled channel effects in ref.~\cite{Zhou}, indicates that the 
interpretation suggested by refs.~\cite{BB1,BB2} may not be valid. 
However, the model of ref.~\cite{Zhou} also fails to reproduce 
the experimentally observed $\pi\pi$ spectrum.

The results obtained when the energy distribution for the decay 
$\Upsilon(3S)\rightarrow \Upsilon\,\pi^+\pi^-$ is 
calculated with the same model as that employed for the 
$\Upsilon'\rightarrow \Upsilon\,\pi^+\pi^-$ decay are shown in 
Fig.~\ref{3Sfig}. The width so obtained is large, 20.1 keV, mainly because 
the $\pi\pi$ momentum $|\vec q\,|$ may extend to over 800 MeV, and the 
form of the distribution does not agree with the experimentally determined 
double-peaked structure. However, if only single quark diagrams are 
considered, the calculated distribution does appear quite similar to 
that obtained in ref.~\cite{Belanger}, where $\pi\pi$ final state 
interactions have been taken into account. In that case the width is 
obtained here as 25.1 keV.

However, it is interesting to see what may be learned by using models of 
the present type to fit the observed double-peaked 
structure of the $\Upsilon(3S)\rightarrow \Upsilon\,\pi^+\pi^-$ 
energy spectrum. In this case there are three parameters 
($\lambda$,~$m_\sigma$ and~$\Gamma_\sigma$) that may be used as free 
parameters. It turns out, that the present model does in fact have 
sufficient freedom to accommodate a double-peaked $\pi\pi$ energy 
distribution, as 
may be seen from Fig.~\ref{3Sfig}. An optimal description of the $\pi\pi$ 
distribution is obtained for the parameter values $m_\sigma$ =
1400~MeV, $\Gamma_\sigma$ = 100 MeV, $\lambda = 2.7\cdot 10^{-3}$, and
gives $\Gamma_{\pi^+\pi^-}$ = 1.07 keV. Since the employed scalar meson 
is now much heavier, the contributions from the pion exchange and single 
quark amplitudes are of equal magnitude, which makes it possible to obtain 
a double-peaked spectrum. This scalar meson mass falls within the range of 
the empirical scalar resonances $f_0$(1370) with mass 
1200-1500 MeV and a width of 200-500 MeV, and $f_0$(1500) with mass 
$1500 \pm 10$ MeV and a width of $112 \pm 10$ MeV~\cite{PDG}. Both of 
these states have, analogously to the $\sigma$, a 
strong coupling to $\pi\pi$. As a relativistic treatment of the pion 
exchange amplitudes will weaken their contribution slightly, the 
$f_0$(1500) appears to be favored by this phenomenological analysis. Since 
most parts of the present model are concerned with the $\sigma\pi\pi$ 
coupling, 
an explanation for the apparent absence of an intermediate $\sigma$ meson 
in case of the decay $\Upsilon(3S)\rightarrow \Upsilon\,\pi^+\pi^-$ may 
possibly be found in the (nonperturbative) hadronization of the gluon pair 
emitted by the heavy constituent quark.

Finally, it is worth noting that the $\pi\pi$ energy distribution for the 
decay $\Upsilon(3S)\rightarrow \Upsilon'\,\pi^+\pi^-$ is not of 
decisive importance since the maximum value of $|\vec q\,|$ is only 
$\sim 170$ MeV for that decay mode. This implies that the shape of the decay 
distribution is mostly determined by phase space alone, and cannot be used 
to discriminate between different theoretical models. 
The experimental data obtained by ref.~\cite{3Sexp} is also rather crude 
for that decay. If the parameters given in Table~\ref{restab} are 
employed, the $\pi\pi$ width for the transition 
$\Upsilon(3S)\rightarrow \Upsilon'\,\pi^+\pi^-$ comes to 
$\sim 0.02$ keV, which is much smaller than the empirically determined 
$0.7\pm 0.2$ keV. This fact was also noted by ref.~\cite{Mannel}, where it 
was found that a much larger coupling constant had to be employed. The 
shape of the $\pi\pi$ distribution is however very insensitive to the 
$\sigma$ parameters, and can be fitted with almost any values.

\newpage

\vspace{1.5cm}
\centerline{\bf Acknowledgments}
\vspace{0.5cm}
We thank Ted Barnes for pointing out the interest in $\pi\pi$ decays of 
heavy quarkonia, and Dmitri Kharzeev for an instructive discussion. TL 
thanks the Waldemar von Frenckell foundation for a fund grant. Research 
supported in part by the Academy of Finland through grant No. 43982.

\end{document}